\documentclass[11pt,a4paper,english]{article}
\usepackage{graphicx}
\usepackage{amssymb,latexsym}
\usepackage{amsmath}
\usepackage{epsf}
\usepackage{pdfsync}
\usepackage{array}
\usepackage{epsfig}
\usepackage[parfill]{parskip}
\usepackage[matrix,arrow,color]{xy}
\usepackage[usenames]{color}
\usepackage{hyperref}
\usepackage{mathrsfs}
\usepackage[font={small}]{caption}
\usepackage{subcaption}
\usepackage[export]{adjustbox}
\usepackage{wrapfig}
\usepackage{microtype}
\usepackage{slashed}
\usepackage{ytableau}
\usepackage{mathtools}
\newdimen\tableauside\tableauside=1.0ex
\newdimen\tableaurule\tableaurule=0.4pt
\newdimen\tableaustep
\def\phantomhrule#1{\hbox{\vbox to0pt{\hrule height\tableaurule width#1\vss}}}
\def\phantomvrule#1{\vbox{\hbox to0pt{\vrule width\tableaurule height#1\hss}}}
\def\sqr{\vbox{%
\phantomhrule\tableaustep
\hbox{\phantomvrule\tableaustep\kern\tableaustep\phantomvrule\tableaustep}%
\hbox{\vbox{\phantomhrule\tableauside}\kern-\tableaurule}}}
\def\squares#1{\hbox{\count0=#1\noindent\loop\sqr
\advance\count0 by-1 \ifnum\count0>0\repeat}}
\def\tableau#1{\vcenter{\offinterlineskip
\tableaustep=\tableauside\advance\tableaustep by-\tableaurule
\kern\normallineskip\hbox
    {\kern\normallineskip\vbox
      {\gettableau#1 0 }%
     \kern\normallineskip\kern\tableaurule}%
  \kern\normallineskip\kern\tableaurule}}
\def\gettableau#1 {\ifnum#1=0\let\next=\null\else
  \squares{#1}\let\next=\gettableau\fi\next}

\input{xy}
\xyoption{all}

\input{epsf}

\makeatletter

\usepackage{setspace}

\usepackage{lscape}

\catcode`@=11
\renewcommand{\section}
{\@startsection{section}{1}{0pt}{\medskipamount}{\medskipamount}{\large\bf}}
\makeatletter\renewcommand{\subsection}
{\@startsection{subsection}{2}{\z@}{-3.25ex plus -1ex minus -.2ex}
{1.5ex plus .2ex}{\it }}

\numberwithin{equation}{section}
\catcode`@=12
\newcommand{\ban}{\begin{eqnarray}}
\newcommand{\ean}{\end{eqnarray}}

\newcommand{\Tr}{{\rm Tr}}

\newcommand{\cN}{{\cal N}}
\newcommand{\cM}{{\cal M}}
\newcommand{\cS}{{\cal S}}

\newcommand{\cH}{{\cal H}}

\newcommand{\cE}{{\cal E}}
\newcommand{\cO}{{\cal O}}

\newcommand{\cT}{{\cal T}}
\newcommand{\cK}{{\cal K}}

\newcommand{\DT}{{\tt DT}}

\newcommand{\HEll}{\mathsf{H\text{-}Ell}}

\newcommand{\complex}{{\mathbb C}} %% complex numbers
\newcommand{\zed}{{\mathbb Z}} %% integers
\newcommand{\real}{{\mathbb R}} %% real numbers
\newcommand{\torus}{{\mathbb T}}

\def\e{{\,\rm e}\,}

\newcommand{\ch}{{\rm ch}}

\def\ii{{\,{\rm i}\,}}
\def\dd{{\rm d}}

\newcommand{\rank}{\mathrm{rank}}
\newcommand{\motive}{\mathbb{L}}

\def\beq{\begin{equation}}
\def\bee{\begin{equation}}
\def\eeq{\end{equation}}
\def\bea{\begin{eqnarray}}
\def\eea{\end{eqnarray}}
\def\bd{\begin{displaymath}}
\def\ed{\end{displaymath}}

\newcommand{\Cint}{\int\kern-10.5pt-\kern7pt}

\newcommand{\PP}{{\mathbb{P}}}

\newcommand{\be}{\begin{equation}}
\newcommand{\ee}{\end{equation}}
\newcommand{\bal}{\begin{align}}
\newcommand{\eal}{\end{align}}

\newcommand\fverbit{\egroup\item[\fbox{\unhbox\pippobox}]}
\newbox\pippobox

{

\def\be{\begin{equation}}
\def\ee{\end{equation}}
\def\bea{\begin{eqnarray}}
\def\eea{\end{eqnarray}}

\newtheorem{lemma}[equation]{Lemma}

\newtheorem{proposition}[equation]{Proposition}

\newcommand{\Proof}[1]{\noindent\underline{\textsf{Proof}}: #1 \hfill
  $\blacksquare$\\}

\begin{document}

\begin{titlepage}
\setcounter{page}{1}

\vskip 5cm

\begin{center}

\vspace*{3cm}

{\Huge Hodge-Elliptic genera, K3 surfaces and \\[9pt] Enumerative Geometry}
%\\[8pt]
%{\Large  -- (Refined) BPS states from $\mathsf{HEll}$ --}

\vspace{15mm}

{\large\bf Michele Cirafici}
\\[6mm]
\noindent{\em Dipartimento di Matematica e Geoscienze, Universit\`a di Trieste, \\ Via A. Valerio 12/1, I-34127 Trieste, Italy, 
\\ Institute for Geometry and Physics (IGAP), via Beirut 2/1, 34151, Trieste, Italy
\\ INFN, Sezione di Trieste, Trieste, Italy 
}\\[4pt] Email: \ {\tt michelecirafici@gmail.com}

\vspace{15mm}

\begin{abstract}
\noindent

K3 surfaces play a prominent role in string theory and algebraic geometry. The properties of their enumerative invariants have important consequences in black hole physics and in number theory. To a K3 surface string theory associates an Elliptic genus, a certain partition function directly related to the theory of Jacobi modular forms. A multiplicative lift of the Elliptic genus produces another modular object, an Igusa cusp form, which is the generating function of BPS invariants of $\mathrm{K3} \times E$. In this note we will discuss a refinement of this chain of ideas. The Elliptic genus can be generalized to the so called Hodge-Elliptic genus which is then related to the counting of refined BPS states of $\mathrm{K3} \times E$. We show how such BPS invariants can be computed explicitly in terms of different versions of the Hodge-Elliptic genus, sometimes in closed form, and discuss some generalizations.

\end{abstract}

\today

\end{center}
\end{titlepage}

%\newpage

\tableofcontents

%\newpage

\section{Introduction}

% don't forget to check numerically the results against wendland

%CLASH OF NOTATIONS: t in K3 geometry and t in rep theory. y is exp z, but z is used in the computation of elliptic genus. in elliptic genus m is baasically the central charge, better notation (it is used after). Variety: X, S and so on? maybe call CY X and generic varieties Y and surfaces S? but Ell is for generic var. Replace $\chi_n$ with $ch (S^nT)$. In ELL and HEll the 1/y is a normalization so it is outside the integral in $\int ch \ todd$. Understand the signs in the generating functions. Go through literature on hell and cite all

String theory compactifications on $\mathrm{K3}$ surfaces underlie a series of connections between gravitational physics, number theory and geometry. The Elliptic genus, a certain partition function in conformal field theory, transforms as a Jacobi modular form \cite{Kawai:1993jk,Witten:1986bf}; its coefficients are related to the irreducible representations of the Mathieu group \cite{Eguchi:2010ej}; its multiplicative lift, a certain Siegel modular form, counts BPS invariants which have the interpretation as black hole microstates in supergravity \cite{Dabholkar:2012nd,Dijkgraaf:1996xw}. See \cite{Anagiannis:2018jqf,Kawai:1997em,Sen:2007qy} for topical reviews. 

In this note we will focus on the connection with enumerative geometry and investigate how it is modified when the Elliptic genus is replaced by the Hodge-Elliptic genus, introduced in \cite{Kachru:2016igs}. In particular we will focus on its relation with certain refined enumerative invariants of $\mathrm{K3} \times E$, with $E$ an elliptic curve, see also \cite{Cheng:2015kha}. The Hodge-Ellitpic genus for K3 is invariant under complex structure deformations, and therefore its multiplicative lift should lead to objects invariant under such deformations. 

The Hodge-Elliptic genus is however not invariant over the K3 moduli space, but jumps at specific jumping loci. The origin of these jumping phenomena, different from wall-crossing, is that the Hodge-Elliptic genus is not an index and is therefore sensitive to the appearance of boson-fermion pairs which would leave an index invariant. Remarkably such an object is often computable in closed form. 

In this note we will concentrate on three specific definitions of such refined partition functions: the generic, the orbifold and the complex Hodge-Elliptic genera. The generic Hodge-Elliptic genus was introduced in \cite{Wendland:2017eiw} and is conjectured to capture physically the refined partition function at a generic point in the moduli space. It can be computed under reasonable assumptions using conformal field theory methods and geometrically reflects the use of the chiral de Rham complex to model twisted sigma models \cite{Kapustin:2005pt}. The orbifold Hodge-Elliptic genus was introduced in \cite{Kachru:2016igs} and captures the physics at an orbifold point. It can be computed in closed form and counts extra states with respect to the generic Hodge-Elliptic genus. The complex Hodge-Elliptic genus was also introduced in \cite{Kachru:2016igs} as a refinement of the usual Elliptic genus, by replacing its geometric formulation as the Euler characteristic of a certain bundle, with the weighted sum of the dimension of certain cohomology groups. Also this partition function contains extra states with respect to the generic Hodge-Elliptic genus. While we couldn't determine precisely the locus in the $\mathrm{K3}$ moduli space where such states appear, it is reasonable to expect that an object invariant under complex structure deformations does indeed occur physically as a refinement of the Elliptic genus at some point of the moduli space. We therefore include it in our analysis, and indeed one of our results is an explicit closed form for the complex Hodge-Elliptic genus. In the process of doing so we also present an explicit computation of the ordinary Elliptic genus, by direct integration from its definition, which highlights the geometric origin of each of its terms. The closed form computation of the complex Hodge-Elliptic genus can be extended to certain twining genera.

The Elliptic genus of a $\mathrm{K3}$ surface determines the enumerative geometry of reduced Donaldson-Thomas invariants of $\mathrm{K3} \times E$. We investigate how this relation is generalized in the case of the refined partition functions. Following \cite{Kachru:2016igs} we conjecture that the resulting generating functions capture a refinement of ordinary Donaldson-Thomas theory. We exhibit closed form formulas for the first terms of the generating functions in all cases, and a prescription to compute higher order terms order by order. Interestingly, the specific form of the generic Hodge-Elliptic genus suggest that the moonshine phenomenon should be visible geometrically in the refined Donaldson-Thomas theory.

Ordinary Donaldson-Thomas theory is blind to the moonshine phenomenon due to a subtle interplay between the coefficients of the Elliptic genus, which we discuss in the text. It has been suggested in \cite{Katz:2014uaa} that the situation could be different in the refined theory. However a closer investigation reveals that this is not the case \cite{Harvey:2017xdt} for the generating function introduced in \cite{Katz:2014uaa}, but leaves the door open for higher order generating functions. Our results seem to imply that this is indeed the case.

This paper is organized as follows. Section \ref{K3surfaces} contains a brief review of certain aspects of the geometry of $\mathrm{K3}$ surfaces, of Elliptic genera and modular forms, and of enumerative geometry. We also connect these aspects explicitly, by showing how the terms in the elliptic genus count certain curves and how this counting is reflected in its (mock) modularity. In Section \ref{HEgenera} we discuss Hodge-Elliptic genera and quickly review the orbifold and generic refined partition sums. We also present an explicit closed form derivation of the complex Hodge-Elliptic genus; in order to do so we revisit the Elliptic genus and compute it explicitly from its integral definition. In doing so we elucidate the geometric interpretation of each term as it arises from the characters of the symmetric product of the tangent bundle to the $\mathrm{K3}$. Finally we show explicitly that each Hodge-Elliptic genus admits a refined decomposition in terms of the holomorphic characters of the superconformal algebra. Section \ref{Enumerative} contains an explicit derivation in closed form of certain terms in a refinement of Donaldson-Thomas theory, predicted using the Hodge-Elliptic genera. Higher order terms can be derived with a similar prescription order by order. Finally Section \ref{Mathieutwined} extends the computation of the complex Hodge-Elliptic genus to certain twined case, obtained by working equivariantly with respect to certain finite order symplectic automorphisms of $\mathrm{K3}$ surfaces. The last Section \ref{Conclusions} summarizes our findings and offers some discussions of open problems.

Two appendices contain some technical results. The supporting \textsc{mathematica} file contains most of the computations hereby discussed \cite{mathfile}.

\section{K3 surfaces, the Elliptic genus and enumerative geometry} \label{K3surfaces}

In this Section we review some aspects of the relation between $\mathrm{K3}$ surfaces and the elliptic genus. We will show how known facts about the Elliptic genus can be put in direct relation with certain Donaldson-Thomas invariants.

\subsection{Some geometric aspects of K3 surfaces}

Here we review some facts about K3 and geometry which will be used along the paper

Recall that for $S$ a K3 surface $H^1 (S , \cO_S) = 0$ and $\Omega^2_S \simeq \cO_s$. The cotangent sheaf $\Omega_S$ is locally free and of rank two and its determinant is trivial $\cK_S \simeq \cO_S$. Furthermore we have the isomorphism $\cT_S \simeq \Omega_S$. The Hodge numbers are given by $h^{p,q} (S) = \dim H^q (S , \Omega^p_S) $. In particular $h^{p,q}=1$ for $(p,q)=(2,0),(0,2),(0,0),(2,2)$, $h^{1,1}= 20$ and all the other Hodge numbers are vanishing. Note that $h^0 (S, \Omega_S) = h^0 (S , \cT_S) = 0$ and there are no global nontrivial vector fields on $S$.

For a rank $r$ sheaf $\cE$ on $S$, the splitting principle allows us to write the Chern polynomial as
\be
c(\cE ; \mathsf{t} ) = \prod_{i=1}^r \left( 1 + a_i \, \mathsf{t}  \right) \, ,
\ee
with $\mathsf{t}$ a formal parameter. The Chern character is given by
\be
\mathrm{ch} (\cE) = r +  c_1 (\cE) +\frac12 \left( c_1 (\cE)^2 - 2 c_2 (\cE) \right) = \sum_{i=1}^r \e^{a_i} \, ,
\ee
and the Todd class by
\be
\mathrm{Todd} (\cE) = 1 + \frac12 c_1 (\cE) + \frac{1}{12} \left( c_1 (\cE)^2 + c_2 (\cE) \right) = \prod_{i=1}^r \frac{a_i}{1 - \e^{-a_i}} \, .
\ee
The Euler characteristic of a sheaf $\cE$ can be computed via the Riemann-Roch theorem
\be
\chi (\cE) = \int \ch (\cE) \, \mathrm{Todd} (\cT) \, ,
\ee
where $\cT$ is the holomorphic tangent space to $S$. This theorem can be used to compute geometrical quantities associated with the $\mathrm{K3}$. For example, from the fact that $\chi (\cO) = 2$ we deduce an expression for the integral of the second Chern class
\be
\chi (\cO) = 2 = \int  \ch (\cO) \, \mathrm{Todd} (\cT) = \frac{1}{12} \int c_2 (\cT) \, .
\ee
The tangent sheaf of a $\mathrm{K3}$ surfaces satisfies
\be
c (\cT ; \mathsf{t} ) = (1 + x \, \mathsf{t}) (1 - x \, \mathsf{t}) = 1 - x^2 \, \mathsf{t}
\ee
and therefore
\be \label{intx24}
\int x^2 = - \int c_2 (\cT) = - 24 \, ,
\ee
a fact that we will use extensively when computing the Elliptic genus.

\subsection{The Elliptic genus}

The definition of the Elliptic genus is rooted in conformal field theory. When the string propagates in a Calabi-Yau variety the worldsheet theory is described by a $\cN=2$ superconformal algebra. This algebra consists of two copies, the left moving and the right moving sectors. If we let $L_0$ and $J_0$ denote respectively the zero-modes of the Virasoro and R-symmetry current generators, the Elliptic genus is defined as a trace over the RR sector
\be \label{EllCFT}
\mathsf{Ell}  (\tau , z) = \Tr_{RR} \left( (-1)^{J_0 + \overline{J}_0} \, y^{J_0} \, q^{L_0 - \frac{c}{24}} \, \overline{q}^{\overline{L}_0 - \frac{\overline{c}}{24}} \right) \, .
\ee
We have introduced the notation $q = \e^{2 \pi \ii \tau}$ and $y = \e^{2 \pi \ii z}$. Due to supersymmetry, this quantity is independent of $\overline{\tau}$ and depends holomorphically on $\tau$ and $z$. Physically the Elliptic genus counts RR states which are in the ground state in the left moving sector and are unconstrained in the right moving sector. It can be understood as computing the dimensions of certain cohomology groups graded by the $L_0$ and $J_0$ quantum numbers. The existence of an inner automorphism of the superconformal algebra, the spectral flow, implies that the Elliptic genus is a weak Jacobi form of weight zero and index $m = \frac{c}{6}$, where $c$ is the central charge. The spectral flow symmetry also implies that Fourier expansion of the Elliptic genus has the structure
\be
\mathsf{Ell}  (\tau , z) = \sum_{n \in \zed_+, \, \ell \in \zed} c \left(n , \ell \right) \, q^n \, y^\ell = \sum_{n \in \zed_+, \, \ell \in \zed} c \left(4 \,m\,  n - \ell^2 \right) \, q^n \, y^\ell
\ee
since the combination of operators which is invariant under spectral flow is precisely $4 \, m \, L_0 - J_0^2$.

In the case of a $\mathrm{K3}$ surface, supersymmetry is enhanced and the Elliptic genus is a weak Jacobi form of weight zero and index 1. The space of these forms is one-dimensional and generated by the form  $\phi_{0,1} (\tau , z)$. Explicitly the Elliptic genus can be written as
%\be
%\mathsf{Ell}  (\tau , z) = 2 \phi_{0,1} (\tau , z) = 8 \sum_{i=2,3,4} \left( \frac{\theta_{i} (\tau , z)}{\theta_i (\tau , 0)}\right) \, .
%\ee
%This expression can also be written as
\be \label{Ell-Jacobi}
\mathsf{Ell}  (\tau , z) = 2 \, \phi_{0,1} (\tau , z)  = \frac{\theta_1 (\tau , z)^2}{\eta (\tau)^3} \left[
24\, \mu (\tau , z) + H (\tau)
\right] \, .
\ee
Here
\begin{align}
\theta_1 (\tau , z) &= \ii q^\frac18 y^{-\frac12} \prod_{n=1}^\infty (1-q^n) (1-q^{n-1} y) (1 - q^n y^{-1}) \, , \\
\eta (\tau) &= q^{\frac{1}{24}} \prod_{n=1}^\infty (1-q^n) \, .
\end{align}
are a theta function and the Dedekind eta function. Furthermore we have introduced the Appell-Lerch sum
\be \label{Appell-Lerch}
\mu (\tau , z) = \frac{-\ii y^{\frac12}}{\theta_1 (\tau, z)} \sum_{\ell = -\infty}^{+ \infty} \frac{(-1)^\ell y^{\ell} \, q^{\ell (\ell + 1) /2}}{1 - y \, q^\ell} \, ,
\ee
and the function
\be \label{FunctionH}
H (\tau) =  \frac{-2 E_2 (\tau) + 48 F_2^{(2)}}{\eta (\tau)^3} = q^{-\frac18} \left( -2 + \sum_{n=1}^{\infty} c_H (n) q^n \right) = 2 q^{-\frac18} \left(  -1 + 45 q + 231 q^2 + 770 q^3 + \cdots
\right) \, ,
\ee
where $E_2 (\tau)$ is the weight two (quasi-modular) Eisenstein series
\be
E_2 (\tau) = 1 - 24 \sum_{n=1}^\infty \frac{n q^n}{1- q^n} \, ,
\ee
and
\be
F_2^{(2)} (\tau) = \sum_{\underset{r > s > 0}{r-s = 1 \mathrm{mod} 2}} (-1)^r \, q^{\frac{r s}{2}} \, .
\ee

The function $H(\tau)$ is a mock modular form with shadow $24 \, \eta (\tau)^3$. This means that 
\be
\widehat{H} (\tau) = H (\tau) + 24 (4 \ii)^{-1/2} \int_{- \overline{\tau}}^\infty (z + \tau)^{-1/2} \, \overline{\eta (- \overline{z})} \dd z  
\ee
is a weight 1/2 modular form for $SL(2 ; \zed)$. Similarly $\mu (\tau, z )$ is a mock modular form with shadow $- \eta (\tau)^3$, so that
\begin{align}
\widehat{\mu} (\tau , z) &= \mu (\tau,z) - (4 \ii)^{-1/2} \int_{- \overline{\tau}}^\infty (z + \tau)^{-1/2} \, \overline{\eta (- \overline{z})} \dd z  % \cr & = \mu (q,y) + \frac{\ii}{2} R (u-v , \tau)
\end{align}
is a weight 1/2 modular form for $SL(2 ; \zed)$. A striking fact about \eqref{Ell-Jacobi} is that the relative coefficients are such that the shadows of the two mock modular forms cancel exactly and the full Elliptic genus is an ordinary Jacobi form. See \cite{Dabholkar:2012nd} for a more complete discussion.

%It is natural to conjecture that the last term transforms as a mock modular form with shadow $- 2 \eta (q)^3$. We will try to show this following Zwegers proof for $\mu$, theorem 1.11
%In Zwegers $\mu (u,v,\tau)$ is ours with $u=v$. So we have the special case
%\be
%R (z=0 , \tau) = \sum_{n \in \frac12 + \zed} \left(
%\mathrm{sgn} (n) - \mathrm{Erf} (n \sqrt{2 \im \tau} ) \right) (-1)^{n-\frac12} q^{-\frac{n^2}{2}}
%\ee
%The modular transformations of this function are known. So it only remains to understand the modular transformations of $h_N (q)$.

Geometrically the Elliptic genus is the holomorphic Euler characteristics of a certain sheaf, constructed out of a sheaf $\cE$ on a variety $Y$. If we denote $r = \rank \, \cE$ and $d = \dim_\complex  Y$, then
\begin{align} \label{defEll}
\mathsf{Ell} (Y ; \tau , z) &= \ii^{r-d} q^{(r-d)/12} \, y^{-r/2} \chi \left( Y ,\mathbb{E} \right) \,  ,
\end{align}
where explicitly
\be \label{bundleEll}
\mathbb{E} =  \bigotimes_{n=1}^{\infty} \, \Lambda_{-y q^{n-1}} \cE \otimes \bigotimes_{n=1}^{\infty} \Lambda_{-y^{-1} q^n} \cE^{\vee} \otimes \bigotimes_{n=1}^{\infty} \cS_{q^n} \cT_Y \otimes \bigotimes_{n=1}^\infty \cS_{q^n} \, \cT^*_Y \, ,
\ee
and we have introduced the notation
\be
\Lambda_q \, \cE = \bigoplus_{k=0}^d \, q^k \, \Lambda^k \, \cE \, , \qquad \cS_q  \, \cE = \bigoplus_{k=0}^\infty q^k \, \cS^k \cE \, .
\ee
We will often drop the explicit dependence on $Y$ from the notation when it is clear from the context. In the following we will take $\cE = \cT_Y$ the holomorphic tangent bundle. In the case of a $\mathrm{K3}$ surface we will simply call this $\cT$. By using the Riemann-Roch theorem the Elliptic genus can be written as
\be
\mathsf{Ell} (Y ; \tau , z) = \frac1y \int \ch(\mathbb{E} ) \, \mathrm{Todd} (\cT) \, .
\ee

The Elliptic genus captures information about certain BPS states in compactifications of the type II string on $\mathrm{K3} \times E$ \cite{Dijkgraaf:1996xw}, where $E$ is an elliptic curve. These are states corresponding to the partition function of the second quantized string theory on $\mathrm{K3} \times E$. The argument of \cite{Dijkgraaf:1996xw} identifies such a partition function with the partition function of a single string which propagates on the symmetric product. The latter is expressed as
\be \label{HilbK3}
\sum_{n=0}^\infty p^n \, \mathsf{Ell} \left( \mathsf{Hilb}^n (\mathrm{K3}) ; \tau , z \right) = p \, \frac{\varphi_{10,1} (\tau, z)}{\Phi_{10} (\sigma , z , \tau)}
\ee 
in terms of the Siegel modular form of weight 10, computed as a multiplicative lift of the Elliptic genus
\be \label{Siegel}
\Phi_{10} (\sigma , z , \tau) = p \, q \, y \, \prod_{(n,m,s) > 0} \, (1 - p^n \, y^s \, q^m)^{c \left(n \, m , s \right)}  \, ,
\ee
where the product is over all $s \in \zed$ and $n,m \ge 0$ such that one of the following two conditions hold
\begin{itemize}
\item $n > 0$ or $m >0$,
\item $n=m=0$ and $s < 0$.
\end{itemize}
Note that the function \eqref{Siegel} is symmetric in the exchange $p \leftrightarrow q$.
Finally the Jacobi form $\varphi_{10,1} (\tau , z)$ which appears in \eqref{HilbK3} can also be thought of as the generating function of $\chi_y$ genera
\be
\sum_{n=0}^\infty q^n \, \chi_{-y} \left( \mathsf{Hilb}^n (\mathrm{K3}) \right) = \prod_{k=1}^\infty \frac{1}{(1 - y \, q^k)^2 \, (1-q^k)^{20} \, (1-q^k y^{-1})^2} \ = q \, \frac{y-2+y^{-1}}{\varphi_{10,1} (\tau ,z )}
\ee 
with
\be
\chi_y (Y) = \sum_{p,q} (-1)^q h^{p,q} (Y) y^p \, .
\ee

\subsection{Donaldson-Thomas theory of $\mathrm{K3} \times E$} \label{DTK3xE}

Donaldson-Thomas theory is the mathematical formalism underlying BPS states formed by bound states of D-branes with electric and magnetic charges. In its simplest version, one considers bound states of a single D6 brane with a gas of D2-D0 branes on a Calabi-Yau $X$. These configurations are parametrized by the Hilbert scheme
\be
\mathrm{Hilb}^{C , n} (X) = \left\{ Z \subset X \, \vert \, [Z] = C \in H_2 (X) , \ n = \chi (\cO_Z) \right\} \, .
\ee
The Donaldson-Thomas invariant is defined via virtual integration over this moduli space
\be
\DT_{C,n} (X) = \int_{[\mathrm{Hilb}^{C , n} (X) ]^{\mathrm{vir}}} \ 1 \ .
\ee
In the case at hand $X = \mathrm{K3} \times E$, with $E$ an elliptic curve. This is the main example that we will consider and therefore we shall often omit the specification of the variety $X$ from the notation. In this case one can parametrize the curve $C$ as $C = \beta + d E$, where $\beta$ is (the push-forward of) a primitive curve in the $\mathrm{K3}$ with $\beta^2 = \langle \beta , \beta \rangle = 2 h-2$, in terms of the intersection pairing. In this case all the relevant invariants vanish; this follows from the localization formulas, since the torus $E$ acts on itself, and therefore on the moduli space, freely. Physically this vanishing is related to the presence of an extra fermionic zero mode $\mathrm{K3} \times E$ compactifications.

To avoid this problem, the relevant moduli space is the quotient of the Hilbert scheme $\mathsf{Hilb}^{C,n} (\mathrm{K3} \times E) / E$. Then one can introduce (reduced) BPS invariants as $\DT_{h,d,n} (X) := \DT_{\beta + d E , n} (X)$ corresponding to the Hilbert scheme $\mathrm{Hilb}^{h,d,n} (X) := \mathrm{Hilb}^{\beta + d E , n} (X) / E$. It is useful to encode this information in the generating function
\be \label{DTgenfun}
\DT (X) = \sum_{h=0}^{\infty} \DT_h (X) p^{h-1} = \sum_{h,d \ge 0 , n\in \zed} \DT_{h,d,n} (X) p^{h-1} q^{d-1} (-y)^n
\ee
where we have introduced the counting parameter  $p = \e^{2 \pi \ii \sigma}$. 

The counting parameters have the following physical interpretation. BPS states on $X$ are parametrized by a certain vector of electric and magnetic charges. However due to T-duality this information is redundant, and only three T-duality invariant scalar combination of the charges are physical, as reviewed for example in \cite{Sen:2007qy}. By an appropriate T-duality rotation we can parametrize the T-duality invariant charges as
\be
Q^2 = q_a C^{ab} q_b = 2 h -2 \qquad P^2 = - 2 q_1 p^0 = -2 d \qquad Q \cdot P =  p^0 q_0 = n
\ee
which are then the parameters which enter in the generating function \eqref{DTgenfun}. Here $p^0 =1$ is the D6 brane charge, $q_0 = n$ the number of D0 branes and the remaining $q_1$ and $q_a$ with $a=2,\dots,23$ are the charges of D2 branes wrapping the elliptic curve or 2-cycles in the K3 respectively. Here $C^{ab}$ is the $(3,19)$ intersection product on $H_2 (\mathrm{K3})$. Note that in the generating function \eqref{DTgenfun} the functions $\DT_h$ have fixed electric charge $Q$, which is different from what one normally does in the physics literature. The geometrical reason for this is that it is easier to count invariants by fixing the curve class in the $\mathrm{K3}$ and summing over the covering of the elliptic curve than viceversa. Ultimately this should be just a matter of conventions since the generating function is symmetric in the exchange $p \leftrightarrow q$.

%
%
%
%The setting is string theory on $K3 \times E$, with $E$ an elliptic curve, and the problem is the counting of certain  BPS states and the associated modular properties. Their generating function has a geometrical description in terms of reduced Donaldson-Thomas invariants $\DT$ associated with the reduced Hilbert scheme $\mathsf{Hilb}^{C,n} (\mathrm{K3} \times E) / E$.
%
%One can assemble the generating function 
%\be
%\DT = \sum_{h=0}^\infty \DT_h (\mathrm{K3} \times E) p^{h-1} = \sum \DT_{h,d,n} p^{h-1} q^{d-1} (-y)^n
%\ee
%In the notation one writes the curve class $C$ as $\beta + d \, E$ where $\beta$ is a primitive curve class of $\mathrm{K3}$ with $\beta^2 = 2 \, h -2$. 
%
%We introduce the variables (for future convenience) 
%\be
%q= \e^{2 \pi \ii \tau} \qquad y = \e^{2 \pi \ii z} \qquad p = \e^{2 \pi \ii \sigma} \qquad u = \e^{2 \pi \ii \nu}
%\ee

The generating function of reduced Donaldson-Thomas invariants is conjecturally given by \cite{Bryan:2015uva}
\be
\DT (\mathrm{K3} \times E) = - \frac{1}{\Phi_{10} (\sigma , z , \tau)}
\ee
in terms of the Siegel modular form \eqref{Siegel}. See also \cite{Bryan:2016bse,Bryan:2016gfy}. The Gromov-Witten formulation of this conjecture \cite{Katz:1999xq,OP} was proven in \cite{Oberdieck:2017zit}. 

In particular Bryan proves \cite{Bryan:2015uva}
\begin{align}
\DT_0 (\mathrm{K3} \times E) &= - \frac{1}{q} \, \frac{y}{(1-y)^2} \ \prod_{m=1}^{\infty} (1-q^m)^{-20} (1-y \, q^m)^{-2} (1-y^{-1} \, q^m)^{-2} = - \frac{1}{\varphi_{10,1} (\tau , z)} \label{BryanDT01} \, ,  \\
\DT_1 (\mathrm{K3} \times E) &=-\frac{24}{q} \prod_{m=1}^\infty (1-q^m)^{-24} \left(
\frac{1}{12} + \frac{y}{(1-y)^2} + \sum_{d=1}^\infty \, \sum_{k \vert d} k (y^k - 2 + y^{-k}) \, q^d 
\right) \, , \label{DT1Bryan}
\end{align}
by a direct toric localization computation. This is quite remarkable since $\mathrm{K3}$ is not a toric manifold. However the relevant moduli space admits a stratification where each stratum is separately toric, even if the torus action does not glue to a globally defined action. These results, as we will see momentarily, give a very good control on the precise geometric interpretation of every BPS state.

%
%\paragraph{Physical aspects} % look at precision counting of black hole microstates
%
%Ultimately move this section in something else. There is a confusion in the charges, the guys corresponding to $h$, which are fixed in the DT expansion, are electric charges and don't enter in $P$, which is how DMZ expand. The only electric charge in $P$ is $q^1$ which is the D2 wrapping the $\torus^2$. So it seems exactly backwards. On the other hand the counting formula is invariant for $P$ and $Q$ exchanges. Fixing $P$ probably has to do with a good supergravity description.
%
%Or maybe it is not so direct, since any way also D4 branes have to wrap the same 2 cycles and therefore are counted by the pair $(h,d)$?
%
%
%CAREFUL: maybe if a lose the supergravity interpretation, I shouldn't talk about single center black holes, but just finite and polar part.

%\paragraph{Comparing $\mathsf{Ell}$ with $\DT$} %preliminary things on DT from attempt recover ell from hohn characters

In particular, since the sum in \eqref{HilbK3} starts from $n=0$ and since $\mathsf{Hilb}^1 (K3)= K3$, we find
\be \label{EllDT}
\mathsf{Ell} (\mathrm{K3} ; \tau , z ) = \frac{\DT_1}{\DT_0} \, .
\ee
We will now use this fact to derive a few non trivial identities.

%From $\mathsf{Ell} = \DT_1 / \DT_0$ we see
%\be
%\mathsf{Ell} = \frac
%{
%-\frac{24}{q} \prod_{m=1}^\infty (1-q^m)^{-24} \left(
%\frac{1}{12} + \frac{y}{(1-y)^2} + \sum_{d=1}^\infty \, \sum_{k \vert d} k (y^k - 2 + y^{-k}) \, q^d \right)
%}
%{
%- \frac{1}{q} \, \frac{y}{(1-y)^2} \ \prod_{m=1}^{\infty} (1-q^m)^{-20} (1-y \, q^m)^{-2} (1-y^{-1} \, q^m)^{-2}  
%}
%\ee
Noting that
\be
\frac{\theta_1^2 (\tau , z)}{\eta^3 (\tau)} = - q^{\frac18} \frac{(1-y)^2}{y} \prod_{m=1}^{\infty} \frac{(1 - y \, q^m)^2 (1-q^m \, \frac{1}{y})^2}{(1-q^n)} \, ,
\ee
we can then rewrite the Elliptic genus as
\begin{align}
\mathsf{Ell} (\tau , z)  = 
\left( \frac{\theta_1^2}{\eta^3} \right) \left( - \frac{24}{\eta^3} \right) \left(
\frac{1}{12} + \frac{y}{(1-y)^2} + \sum_{d=1}^\infty \, \sum_{k \vert d} k (y^k - 2 + y^{-k}) \, q^d \right)
=\left( \frac{\theta_1^2}{\eta^3} \right) \left( - \frac{24}{\eta^3} \right) \wp (\tau , y)
\end{align}
dove $ \wp (\tau , y)$ is the Weierstrass $\wp$-function. Let us consider this expression term by term and compare it with \eqref{Ell-Jacobi}. 
%Consider first
%\be
%-\frac{24}{\eta^3} \left(
%\frac{1}{12} + \sum_{d=1}^\infty \, \sum_{k \vert d} k (- 2 ) \, q^d \right)
%\ee
It is easy to see that
\be
\sum_{d=1}^\infty \, \sum_{k \vert d} k (- 2 ) \, q^d = \sum_{m=1}^\infty \sum_{k=1}^\infty \, k (- 2 ) \, q^{m \, k} = -2 \sum_{n=1}^\infty \frac{n \, q^n}{(1-q^n)} 
\ee
and therefore
\begin{align} \label{diag12}
-\frac{24}{\eta^3 (\tau)} \left(
\frac{1}{12} + \sum_{d=1}^\infty \, \sum_{k \vert d} k (- 2 ) \, q^d \right) = -\frac{24}{\eta^3 (\tau)} \left(
\frac{1}{12}  -2 \sum_{n=1}^\infty \frac{n q^n}{(1-q^n)}  \right)
= - 2 \frac{E_2 (\tau)}{\eta^3 (\tau)} \, .
\end{align}
By comparing with the know expression of the Elliptic genus we deduce the fairly non trivial identity
\be \label{vertmu}
- \frac{24}{\eta^3 (\tau)} \left(
 \frac{y}{(1-y)^2} + \sum_{d=1}^\infty \, \sum_{k \vert d} k (y^k + y^{-k}) \, q^d \right) = 24 \, \mu (q , y) + 24 \, \frac{2 F_{2}^{(2)} (\tau)}{\eta^3 (\tau)}
\ee
which we will use in the following. Note that for this identity to hold it is crucial that the $2 F_{2}^{(2)} (\tau) / \eta^3 (\tau)$ term cancels the $y$-independent terms in the Appell-Lerch sum \eqref{Appell-Lerch}, since in the left hand side every term in the $q$-expansion has a $y$-dependent coefficient. The above relation can also be proven indirectly using modularity, by using the fact that the ratio of Jacobi forms $\varphi_{0,1} / \varphi_{-2,1}$ is proportional to the Weierstrass $\wp$-function, see \cite{Dabholkar:2012nd} for example.

The function $H(\tau)$ \eqref{FunctionH}, whose coefficient encode the dimensions of certain irreducible representations of the Mathieu group $\mathbb{M}_{24}$, does not appear to have an independent geometrical interpretation. This fact underlies the difficulty in finding a geometric interpretation of the Mathieu group representations. We will discuss more aspects of the geometry of curves in the target space $\mathrm{K3} \times E$ momentarily.

\subsection{Enumerative aspects of Jacobi forms} \label{aspects}

We will now give an example of what kind of information we can gain from the explicit expression of the enumerative invariants \eqref{BryanDT01}. It was shown in \cite{Dabholkar:2012nd}
 that the coefficients of the expansion of $1 / \Phi_{10}$ at fixed magnetic charges are meromorphic Jacobi forms. These can be decomposed into a finite part, a certain finite linear combination of classical theta series where the coefficients are mock modular forms, and a polar part, that is completely determined by the poles of the original meromorphic Jacobi form. Physically the finite part counts microstates associated with immortal black holes, defined over all the moduli space, while the polar part captures the degeneracies associated with two centered black holes. While this decomposition is perfectly clear from the point of view of modular forms, it begs for a geometric explanation. For example one can ask what is the enumerative content of single-centered black holes, or if we can give a precise mapping between enumerative invariants and black hole microstates. In other words, given a certain Donaldson-Thomas invariant, possibly in terms of a moduli space of schemes, how can we know in purely geometrical terms if it corresponds to a single center black hole or not?
  
In our case the situation is slightly different, since we are considering the coefficients of the expansion of $1 / \Phi_{10}$ at fixed electric charges; nevertheless $\Phi_{10}$ is symmetric under this exchange and the results of \cite{Dabholkar:2012nd} formally still hold. Such an expansion is somewhat unnatural from the physical point of view and possibly needs a clearer interpretation (possibly it holds in a different region of the moduli space). Nevertheless we will show explicitly such decomposition for $\DT_1$.

Before addressing the issue, let us clarify the geometrical content of the generating functions \eqref{BryanDT01}, following \cite{Bryan:2015uva}. Consider $X = K3 \times E$ and let $p_{1,2}$ be the projections onto the first and second factors. Then a curve $C$ is called \textit{vertical} if its projection $p_2 : C \longrightarrow E$ has degree zero, so that the curve lies in $K3 \times pt$. Similarly a curve $C$ is called \textit{horizontal} if the projection $p_1 : C \longrightarrow K3$ has degree zero. A curve which is neither vertical nor horizontal is called \textit{diagonal}. This classification determines a decomposition
\be \label{Hilb-decomp}
\mathrm{Hilb}^{h,d,n} (X) = \mathrm{Hilb}^{h,d,n}_{\mathrm{vert}} (X) \times \mathrm{Hilb}^{h,d,n}_{\mathrm{diag}} (X) \, .
\ee
This decomposition follows from the fact that since $\beta$ is irreducible, any subscheme $Z$ will have a unique component which is either diagonal or vertical, while all the other components are horizontal. In particular if $h=0$ only vertical components are possible.

Let us now focus on the $\DT_1$ contribution. Consider the vertical part in the decomposition \eqref{Hilb-decomp}. In the case $h=1$, we have $\beta^2 = 0$. Equivalently the K3 are elliptically fibered and $\beta$ is the class of the fiber. The fibration $S \longrightarrow \PP^1$ has 24 singular fibers. As proven in \cite{Bryan:2015uva} the computation of the BPS invariants receives contributions from subschemes whose unique vertical component is a smooth fiber:
\be \label{vert-smooth}
-22 \prod_{m=1}^\infty \left( 1 - q^m \right)^{-24} \, ,
\ee
where $22$ is the Euler characteristics of the base $\PP^1$ minus 24 points, and from subschemes whose unique vertical component is a nodal fiber:
\be \label{vert-nodal}
24 \prod_{m=1}^\infty \left( 1 - q^m \right)^{-24} \left[ 1 + \frac{y}{(1-y)^2} + \sum_{d=1}^\infty \, \sum_{k \vert d} k (y^k + y^{-k}) \, q^d \right] \, .
\ee
To these one must add the contribution from the diagonal curves
\be \label{diag}
24 \prod_{m=1}^\infty \left( 1 - q^m \right)^{-24}  \sum_{d=1}^\infty \, \sum_{k \vert d} (-2 k) \, q^d
\ee
These three terms give a concrete geometrical interpretation of \eqref{DT1Bryan}. We will now show how precisely these terms contribute to the black hole partition function.
%We will now see that in the generating functions  the diagonal part together with the smooth fiber part and the nodal fiber part without points form the $E_2 / \eta^3$ contribution to the black hole partition function. 
It follows from \cite{Dabholkar:2012nd}  that we expect that each $\DT_h$ splits into a polar part
\be
\DT_h^{P} = \frac{p_{24} (h+1)}{\eta^{24} (\tau)} \sum_{s \in \zed} \frac{q^{h s^2 + s} \, y^{2 h s + 1}}{(1 - q^h \, y)^2} \, ,
\ee
and a finite part $\DT_h^{F} = \DT_h - \DT_h^{(P)}$. Indeed in our case
\be
\DT_1 = \DT_0 \, \mathsf{Ell} = - 2 \frac{\varphi_{0,1}}{\varphi_{10,1}} = - 2 \frac{\varphi_{0,1}}{\varphi_{-2,1}} \frac{1}{\eta (\tau)^{24}} \, .
\ee
Then we have (from \cite{Dabholkar:2012nd} eq 8.54; both $\varphi_{0,1}$ and $\varphi_{-2,1}$ here are defined with the opposite sign as there) 
\be
\frac{\varphi_{0,1}}{\varphi_{-2,1}} = 12 \mathrm{Av}^{(0)} \left[ \frac{y}{(1-y)^2} \right] + E_2 (\tau) \, ,
\ee
where
\be
\mathrm{Av}^{(m)} [f (y)] = \sum_k q^{m k+2} \, y^{2 k m} f (q^k y)
\ee
is the averaging operator. In our case it is easy to see that
\be
\mathrm{Av}^{(0)} \left[ \frac{y}{(1-y)^2} \right] = \sum_{s \in \zed} \frac{q^s y}{(1 - q^s y)^2} = \frac{y}{(y-1)^2} + \sum_{d=1}^\infty \sum_{k \vert d} k \left( y^k + y^{-k} \right) \, q^d
\ee
We conclude that
\begin{align}
\DT_1^{(P)} &= - \frac{24}{\eta (\tau)^{24}} \, \mathrm{Av}^{(0)} \left[ \frac{y}{(1-y)^2} \right] \\
\DT_1^{(F)} &= - \frac{2}{\eta (\tau)^{24}} \, E_2 (\tau)
\end{align}
Now we have an explicit enumerative interpretation of the Jacobi form $\DT_1$: the BPS states counted by the Fourier part correspond to the diagonal curves \eqref{diag} as well as those curves whose vertical component is a smooth fiber \eqref{vert-smooth} or a nodal curve without points, that is the $y^0$ term in \eqref{vert-nodal}. The rest of the vertical curves in \eqref{vert-nodal} are associated with the polar part. 

This example shows that by carefully analyzing the interplay between modularity and geometry we can give a very explicit description of the BPS states. 

%
%Now the enumerative interpretation of black hole microstates follows. The single center microstates correspond to the diagonal curves as well as those  curves whose vertical component is a smooth fiber and the nodal curves without points (that is the $y^0$ termn) . The rest of the vertical curves, that is excluding the $\frac{1}{12}$ factor in \eqref{boh}, are associated with two center black hole states. 
%
%Indeed the $\frac{1}{12}$ factor in \eqref{} comes from the contribution of vertical curves, with smooth fibers ($Hilb_F$) and curves with nodel fibers ($Hilb_N$) but no points (independent of $y$).
%
%REPLACE fiber with subscheme supported on locus with a component with a vertical curve, or just more precise
%
%ADD DMZ physical interpretation around eq 11.20? can one match the enumerative curves more precisely? maybe one can give the interpretation of each single term in the sum over d and d | k ? recall that sum is equal to
%$$
%\sum_{k=1}^\infty \sum_{d=1}^\infty k (y^k + y^{-k }) q^{m k}
%$$
%which should be the expansion of the average 
%
%
%
%IT IS A BIT SUSPICIOUS THAT WE GET THE SAME RESULT from the p and q expansion? maybe they literally are the same and one can chose a duality frame where electric and magnetic charges are exchanged.
%
%ALSO REMEMBEr only the finite part with positive discriminant correspond to single center black holes. See DMZ sect 11
%

\section{Hodge-Elliptic genera on K3 surfaces} \label{HEgenera}

In this Section we introduce various Hodge-Elliptic genera for K3 surfaces, following \cite{Kachru:2016igs,Wendland:2017eiw}, and discuss some of their properties. The Hodge-Elliptic genus is invariant under complex structure deformations but depends sensitively on the K\"ahler structure. The orbifold Hodge-Elliptic genus \cite{Kachru:2016igs} corresponds to $\mathrm{K3}$ surfaces which are resolutions of a certain quotient, while the generic conformal field theoretic Hodge-Elliptic genus \cite{Wendland:2017eiw} is conjectured to capture the large radius conformal field theory. Both are explicitly computable. We also discuss the complex Hodge-Elliptic genus \cite{Kachru:2016igs} and express it in closed form by using representation theory arguments. In passing we also clarify the structure of each contribution to the Elliptic genus from bundles on $\mathrm{K3}$.

\subsection{Refined partition functions}

Counting functions of BPS states can usually be refined by keeping track of additional quantum numbers. For example the function $1/ \varphi_{10,1}$  can be interpreted as counting 1/2-BPS states in type II string compactifications on $\mathrm{K3} \times E$ with fixed electric charge and transforming in a certain spin representation of $\mathrm{SU}(2)_L$. The function $\varphi_{10,1}$ admits a refinement which was derived by Katz, Klemm and Pandharipande \cite{Katz:2014uaa}
\begin{align}
\phi_{KKP}  (\tau , z , \nu)=& \, q \, (y^\frac12 u^\frac12 - y^{-\frac12} u^{-\frac12}) (y^{-\frac12} u^\frac12 - y^{\frac12} u^{-\frac12})  
\cr & \prod_{m=1} (1-q^m)^{20} (1-q^m u\, y) (1-q^m u \, y^{-1}) (1-q^m y \, u^{-1}) (1 - q^m u^{-1} y^{-1}) \, ,
\end{align}
where we have introduced the new counting parameter $u = \e^{2 \pi \ii \nu}$. Physically $1 / \phi_{KKP}  (\tau , z , \nu)$ refines the counting of $1/ \varphi_{10,1}$  in the sense that the fugacity $u$ keeps track of the spin representation now in $\mathrm{SU}(2)_R$. This function has also a geometrical interpretation as
\begin{align}
\sum_{n \ge 0} & \chi_{\mathrm{Hodge}} \left( \mathsf{Hilb}^n \mathrm{K3} \right) q^{n} = \frac{\, q \, (u-y-y^{-1}+u^{-1})}{\phi_{KKP} (\tau, z , \nu)} \cr
= & \prod_{k=1}^{\infty} (1-q^k)^{-20} (1-u \,y \,q^k)^{-1} (1-u^{-1} y\, q^k)^{-1} (1- u \,y^{-1}q^k)^{-1} (1-u^{-1} y^{-1}q^k)^{-1} 
%& =\frac{p (u-y-y^{-1}+u^{-1})}{\phi_{KKP (\sigma, z , \nu)}}
\end{align}
where the Hodge polynomial for a variety $Y$ is given by
\be
\chi_{hodge} (Y) = u^{-d/2} \, y^{-d/2} \sum_{p,q} \, (-u)^q \, (-y)^p \, h^{p,q} (Y)
\ee
For a $\mathrm{K3}$ surface, $\chi_{\mathrm{Hodge}} (\mathrm{K3}) = 1/(u y) +y/u + 20 + u/y + y \, u$, which reduces to $\chi_{y} (\mathrm{K3}) = 2 \, y+20+2/y$ for $u=1$.

%Up to normalization we can argue
%\be
%\chi_{hodge} (M) = \sum_{p,q} \, (u)^q \, (y)^p \, h^{p,q} (M)
%\ee
%so that by Gottsche formula
%\be
%\sum_{n=0}^\infty \chi_{hodge} (Hilb^n K3) \left( \frac{q}{u y}\right)^n = \prod_{n=1}^\infty  \prod_{i,j}^2 \left( 1 - (-1)^{i+j} \, u^{i-1} y^{j-1} q^n \right)^{-(-1)^{i+j} h^{i,j}}
%\ee
%which reproduces the above result (cfr bakker's thesis).

%Note that (we denote by $\cS$ the symmetrization, or plethistic exponential)
%\be
%\sum_{n \ge 0} \chi_{hodge} \left( \mathsf{Hilb}^n \mathrm{K3} \right) q^{n} = \cS^\bullet \frac{q}{1-q} \chi_{hodge} (K3)
%\ee
%with $\chi_{hodge} (K3) = 1/(u y) +y/u + 20 + u/y + yu$. Similarly
%\be
%\sum_{n \ge 0}  \chi_{-y} \left( \mathsf{Hilb}^n \mathrm{K3} \right) q^{n} = \cS^\bullet \frac{q}{1-q} \chi_{-y} (K3)
%\ee
%with $\chi_{-y} (K3) = 2 y+20+2/y$.

Precisely as the generating function $\varphi_{10,1}$ admits the refinement $\phi_{KKP}$, also the elliptic genus and the associated generating functions can be refined. Kachru and Tripathy \cite{Kachru:2016igs} introduce the conformal field theory Hodge-Elliptic genus as a refinement of the Elliptic genus
\be \label{CFTHEll}
\HEll (\tau , z , \nu) = \Tr \left( (-1)^{J_0 + \overline{J}_0} \, y^{J_0} u^{\overline{J}_0} q^{L_0 - \frac{c}{24}} \right) = \sum_{n,s,m} \, c \left(n,\ell,m \right) q^n y^\ell u^m \, .
\ee
The trace is taken only over states which are arbitrary in the left-moving sector but in a Ramond ground state in the right-moving sector. Such a quantity takes its name from the fact that its first term in the $q$ expansion is the Hodge polynomial of the $\mathrm{K3}$. It is however not a genus in the strict mathematical sense (for example it is non vanishing on the torus). This quantity is moduli dependent and jumps over the Calabi-Yau moduli space, for example at those points where the left chiral CFT algebra is enhanced.

This quantity can also be given a geometric definition on a variety $Y$ using the geometric realization of the Elliptic genus as the Euler characteristic of the sheaf \eqref{bundleEll} \cite{Kachru:2016igs}:
\begin{align} \label{defHEll}
\HEll^c (Y ; \tau , z , \nu) &= \ii^{r-d} q^{(r-d)/12} \, y^{-r/2} \, u^{-d/2} \\ \nonumber
& \sum_{j=0}^r (-u)^j \dim H^j \left( Y , \bigotimes_{n=1}^{\infty} \, \Lambda_{-y q^{n-1}} \cE \otimes \bigotimes_{n=1}^{\infty} \Lambda_{-y^{-1} q^n} \cE^{\vee} \otimes \bigotimes_{n=1}^{\infty} \cS_{q^n} \cT_X \otimes \bigotimes_{n=1}^\infty \cS_{q^n} \, \cT^*_X \right) \, .
\end{align}
As for the Elliptic genus we will only consider the case where $Y$ is a $\mathrm{K3}$ surface and $\cE = \cT$ its holomorphic tangent bundle. We will refer to this as the complex Hodge-Elliptic genus. In general a direct computation of the Hodge-Elliptic genus will give different results when carried out at different points of the moduli space, as we will discuss momentarily. Remarkably in certain cases the Hodge-Elliptic genus is known in closed form. A direct computation of \eqref{CFTHEll} was done in \cite{Kachru:2016igs} at a certain orbifold point and in \cite{Wendland:2017eiw} in the strict large radius limit. We will refer to these as $\HEll^{\mathrm{orb}}$ and $\HEll^g$; when writing $\HEll$ without any further specification we refer to properties which hold for each of the Hodge-Elliptic genera described so  far.  

The main interest of this paper is how Hodge-Elliptic genera are related to enumerative geometry. As shown in \cite{Kachru:2016igs}, essentially following the argument of \cite{Dijkgraaf:1996xw} one can define a generating function
\be \label{liftHEll}
\sum_{k=0}^{\infty} p^k \, \HEll \left( \mathrm{Hilb}^{[k]} (\mathrm{K3}) \right) = \prod_{r>0,s\ge0,t,v} \frac{1}{(1 - q^r y^t p^r u^v)^{c(r\, s,t,v)}} = \frac{p \, \phi_{KKP} (\tau , z , \nu)}{\Phi^{\mathrm{ref}} (\tau , z , \nu, \sigma) } \, .
%\sum_{n=0}^\infty p^n \, \HEll (\mathsf{Hilb^n} (\mathrm{K3})) = \prod_{n >0 , m \ge 0, s, v } \, \frac{1}{(1-q^n \, y^s \, p^m \, u^v )^{c(n m, s, v)}} = p \frac{\phi_{KKP} (\tau, z, \nu)}{\Phi^{\mathrm{ref}} (\tau , z , \nu, \sigma)} \, .
\ee
%The function 
%\be
%\Phi^{\mathrm{ref}} (\tau , z , \nu, \sigma) & = p\, q \, y \prod_{(s,t,r,u) > 0} \left( 1 - q^s y^t p^r u^v \right)^{c (r\, s ,  t  , v)}
%\ee
%is defined in terms of the coefficients of the Hodge-Elliptic genus and plays the role of a refined version of the Siegel modular form $\Phi_{10}$ we encountered in \eqref{Siegel}. The product is over all the $s, v \in \zed$ and $n,m \ge 0$ such that one of the following two conditions hold
%\begin{itemize}
%\item $n > 0$ or $m >0$
%\item $n=m=0$ and $s < 0$.
%\end{itemize}
%Note that $\Phi (\sigma, z, \tau, \nu) $ is symmetric in the exchange $q \longrightarrow p$, as $\Phi_{10}$ is.

The authors of \cite{Kachru:2016igs} conjecture that this $\Phi^{\mathrm{ref}} (\tau , z , \nu, \sigma)$ is the generating function of motivic/refined Donaldson-Thomas invariants. Since it is defined in terms of the Hodge-Elliptic genus, its precise form will depend on the Calabi-Yau moduli. In this note we will study various $\Phi^{\mathrm{ref}} (\tau , z , \nu, \sigma) $ corresponding to different version of the Hodge-Elliptic genus and use it to make prediction for enumerative invariants. These matters will be discussed in Section \ref{Enumerative}

\subsection{BPS jumping loci}

The flavoured partition function \eqref{CFTHEll} is not an index and can exhibit jumping behaviour at certain points in the moduli space \cite{Kachru:2017yda,Kachru:2017zur}. This behaviour is generically different from wall-crossing and is due to the unprotected nature of the new partition function. For example the indexed count can differ from the flavoured partition function at generic points in the moduli space due to the presence of boson-fermion pairs (where the statistics refers to the fermion number appearing in the index, so typically we really are talking about different multiplets) whose contribution to the Elliptic genus cancel exactly. Additionally an extra chiral current can appear at special points in the moduli space, leading to more state appearing (always in bose-fermi pairs to leave the indexed count invariant). A physical interpretation could be that for special loci in the moduli space some particles in the full spectrum are now annihilated by a certain supercharge (or a combination thereof). Another difference with wall-crossing is that jumping loci are typically  of higher co-dimension.

For example the moduli space of type IIA compactification on K3 surfaces has the form of a locally symmetric space
\be
\cM (p,q) = O(p,q ; \zed) \backslash O(p,q ; \real) / \left( O(p) \times O(q) \right) 
\ee
which can be understood as the moduli space of lattices $\Gamma^{p,q}$ of signature $(p,q)$. By adopting this perspective the appearance of extra chiral currents corresponds to loci in the moduli space where the lattice generated by a collection of $k$ vectors becomes purely left moving, corresponding to subvarieties of the form $\cM (p,q-k) \subset \cM (p,q)$ and called \textit{special cycles}. In algebraic geometry such loci where the rank of the lattice changes abruptly are known as Noether-Lefschetz loci, or generalization thereof, and are often Shimura varieties. Remarkably the formal generating functions of these loci, that is sums whose coefficients are special cycles, are in certain cases (mock) modular forms valued in $H^\bullet (\cM (p,q))$ \cite{Kachru:2017zur}.
 
\subsection{The orbifold Hodge-Elliptic genus}

In \cite{Kachru:2016igs} the Hodge-Elliptic genus is computed at a certain point in the moduli space where the $\mathrm{K3}$ is a resolution of the quotient of $\torus^4$ (a product of two square tori with unit volume) by the $\zed_2$ inversion. Using CFT techniques one finds explicitly\footnote{The sign difference with \cite{Kachru:2016igs} is due to a different convention in the definition of $\theta_1 (\tau , z)$.}
\begin{align}
\HEll^{\rm orb} (K3) =  &8 \Big[- \left( \frac{\theta_1 (\tau , z)}{\theta_1^* (\tau , 0)}  \frac{u^{-1/2}-u^{1/2}}{2}  \right)^2 +
 \left( \frac{\theta_2 (\tau , z)}{\theta_2 (\tau , 0)}  \frac{u^{-1/2}+u^{1/2}}{2}  \right)^2  \cr & +
  \left( \frac{\theta_3 (\tau , z)}{\theta_3 (\tau , 0)} \right)^2 +  \left( \frac{\theta_4 (\tau , z)}{\theta_4 (\tau , 0)} \right)^2
\Big]
\end{align}
with $\theta_1^* (\tau  , 0 ) = - 2 q^{1/8} \prod_{n=1}^\infty (1-q^n)^3$. 
% the difference with kachru is that they don't have an i in theta1
We can rewrite this as
\begin{align}
\HEll^{\rm orb} (K3) &= 2 \left( \frac{1}{u} + u - 2 \right) \left( - \left( \frac{\theta_1 (\tau , z)}{\theta_1^* (\tau , 0)} \right)^2 +
 \left( \frac{\theta_2 (\tau , z)}{\theta_2 (\tau , 0)} \right)^2  \right) + 2 \, \varphi_{0,1} (\tau , z) \cr
 & = 2 \left( \frac{1}{u} + u - 2 \right) \left(  \frac14  \varphi_{-2,1} (\tau , z) +
 \left( \frac{\theta_2 (\tau , z)}{\theta_2 (\tau , 0)} \right)^2  \right) + 2 \, \varphi_{0,1} (\tau , z)
\end{align}
%\textbf{THE SIGN OF THE THETA1 is wrong} because of different conventions with kachru.
Define the function\footnote{Here $M_2 (\Gamma_0 (N))$ denotes the space of weight 2 modular forms on $\Gamma_0 (N)$, the congruence subgroup of the modular group of level $N$. Also recall that $\Gamma (N) \subset \Gamma_1 (N) \subset \Gamma_0 (N) \subset \mathrm{SL} (2 ; \zed)$ and $\Gamma (N') \subset \Gamma (N)$ whenever $N \vert N'$. If $\Gamma \subset \Gamma'$ then $M_k (\Gamma') \subset M_k (\Gamma)$.}
$\Lambda_N (\tau) \in M_2 (\Gamma_0 (N))$ as
\be \label{Lambdadef}
\Lambda_N (q) = N \frac{\dd}{\dd q} \log (\frac{\eta (q^N)}{\eta(q)}) = \frac{N}{24} \left( N E_2 (N \tau)-E_2 (\tau) \right)
\ee
Then using the identity \cite{Duncan:2015xoa}
\be
\left( \frac{\theta_2 (\tau , z)}{\theta_2 (\tau , 0)} \right)^2 = \frac{1}{12} \, \varphi_{0,1} (\tau , z) +2  \, \Lambda_2 (\tau) \, \varphi_{-2,1} (\tau , z)
\ee
one can write
\begin{align}
\HEll^{\rm orb} (K3) %& = 2 \left(\frac{1}{u} + u - 2 \right) \left(\frac14 + 2 \, \Lambda_2 (\tau) \right) \phi_{-2,1} (\tau, z)+ \left( 2 + \frac16 \left( \frac{1}{u} + u - 2 \right) \right) \phi_{0,1} (\tau ,z) \cr
&= \left( \frac53 + \frac{1}{6 u} + \frac{u}{6} \right) \varphi_{0,1} (\tau ,z)  - \left(1 - \frac{1}{2 u} - \frac{u}{2} \right) \left(1 + 8  \Lambda_2 (\tau)  \right) \varphi_{-2,1} (\tau, z) \\
\label{HEllorb}
%\HEll^{\rm orb} (K3) 
 &=\frac{1}{24} \left(\frac{2}{u} + 20 + 2 u \right) \mathsf{Ell} (\tau , z) + \left(1 - \frac{1}{2 u} - \frac{u}{2} \right) \frac{\theta_1 (\tau , z)^2}{\eta (\tau)^6} \left(1 + 8  \Lambda_2 (\tau)  \right)
\end{align}
which has the structure of the sum of two Jacobi forms with coefficients which are $u$-dependent and a weight 2 modular form on $\Gamma_0 (2)$. 

%
%-----------------
%
%SHOULD be completely rechecked in particular the conventions of kachru and the sign of phi-21
%COMPARE THE ABOVE WITH HELL there are quite few similarities, it looks the same except $E_2 (q^2)$ is truncated? Note that E2.... compare with Hell
%
%-----------------

\subsection{The chiral Hodge-Elliptic genus}

The generic conformal field theory Hodge-Elliptic genus corresponds to the partition function \eqref{CFTHEll} computed in the infinite volume limit. It was computed in \cite{Wendland:2017eiw} by a careful analysis of the space of ground states and of the representation theory of the superconformal algebra. The result is
\be \label{genericHEll}
\HEll^g (\tau , z , \nu) = \mathsf{Ell} (\tau , z ) + \left( 2 - \frac{1}{u} - u \right) \frac{\theta_1 (\tau , z)^2}{\eta(\tau)^3} \left[ q^{-1/8} - 2 \mu (\tau , z) \right]
\ee
where the second term on the right hand side is proportional to the character $\mathrm{ch}_{\frac14 , \frac12} (\tau , z)$ of the superconformal algebra, which will however appear later on. This implies that the generic Hodge-Elliptic genus has modular properties which differ from the ones of the Elliptic genus $\mathsf{Ell} (\tau , z)$ essentially by the addiction of the mock modular form $\mu (\tau , z)$.

Remarkably \eqref{genericHEll} has a geometric interpretation in terms of the chiral de Rham complex $\Omega^{ch}$ of $\mathrm{K3}$ introduces in \cite{chiraldR}. Recall that the chiral de Rham complex is a sheaf of vertex operator algebras obtained by gluing together local $(b c - \beta \gamma)$-systems. Taking the sheaf homology $H^\bullet (\mathrm{K3} , \Omega^{ch})$ provides a model for the sigma model Hilbert space of states. Then it is shown in \cite{Wendland:2017eiw} that
\be
\HEll^g (\tau , z , \nu) = (y \, u)^{-1} \sum_{j=0}^2 (- u )^j \, \Tr_{H^j (\mathrm{K3} , \Omega^{ch})}  \left( (-1)^{J_0} y^{J_0} q^{L_0 - \frac12 J_0} \right)
\ee
where the combination $L_0 - \frac12 J_0$ signals that the chiral de Rham complex carries the action of a topologically twisted superconformal algebra.  

By using \eqref{Ell-Jacobi} we can write \eqref{genericHEll} in a form similar to \eqref{HEllorb}
\begin{align} \label{genericHEll2}
\HEll^{g} (\tau , z, \nu) 
%&= \mathsf{Ell} (\tau , z) + \left( 2 - \frac{1}{u} - u \right) \frac{\theta_1 (\tau , z)^2}{\eta (\tau)^3} \left[ q^{-\frac18} - 2 \mu (\tau , z) \right] 
%\cr &
=  \frac{1}{24} \left( 20 + \frac{2}{u} + 2 u \right) \mathsf{Ell} (\tau , z) + \left( 2 - \frac{1}{u} - u \right) \frac{\theta_1 (\tau , z)^2}{\eta (\tau)^3} \left[ \frac{1}{12} H (\tau) + q^{-\frac18} \right] \, .
\end{align}

\subsection{The complex Hodge-Elliptic genus and representation theory.}

We will now outline a procedure to compute \eqref{defHEll} directly from the definition, as an expansion in $q$. The idea is to reduce the computation to a sum of factors which can be read of from the ordinary elliptic genus expansion, with different weights.

% See for example huybrechts.

A $\mathrm{K3}$ surface has strict $\mathrm{SU(2)}$ holonomy and therefore $\mathrm{SU(2)}$ acts on the tangent space $\cT$. By using the splitting principle we can write the character $\ch (\cT) = t + 1/t$ which is then identified with the character of the fundamental representation of SU(2). In the case of SU(2) characters one can compute explicitly the generating function
\be
\sum_{n=0} s^n \chi_n (t) = \frac{1}{(1-s t) (1-\frac{s}{t})} \, .
\ee
We can think of $t$ as a one dimensional module under the action of the diagonal $\mathrm{U}(1) \subset \mathrm{SU}(2)$. Here $\chi_n (t) = \frac{t^{n+1} - t^{-n-1}}{t-t^{-1}} = \sum_{i=0}^n t^{2 i-n}$.

Now we look explicitly at the bundle $\mathbb{E}$ in \eqref{bundleEll}. It has a form of a direct sum of bundles whose coefficients are weighted by $q^n$. The antisymmetric factors only contain a finite number of terms. We can then write
\begin{align}
\mathbb{E} = \bigotimes_{n=1}^{\infty} 
\left(\cO - y q^{n-1} \cT + y^2 q^{2 (n-1)} \cO \right)  \left(\cO - \frac{q^n}{y} \cT + \frac{q^{2 n}}{y^{2}} \cO \right) \left(\bigoplus_{k=0} q^{n k} \cS^k (\cT) \right)  \left(\bigoplus_{l=0} q^{n l} \cS^l (\cT) \right)  \, .
\end{align}
Under the SU(2) action $\cS^k (\cT)$ corresponds to the character $\chi_k$.  The reason why it is useful to think in this terms is that $H^0 (X , \cS^m \cT) = 0$ $\forall m >0$, by a classic result of Kobayashi. Therefore to compute the Hodge-Elliptic genus it is sufficient to single out the terms which admit global sections. The terms which do not admit global sections have trivial $H^0$ and therefore by Serre's duality trivial $H^2$. The strategy of the computation is to use the representation theory of SU(2) to decompose a generic term $\cT^k$ into terms which admit global sections (that are given by $\cO$ and correspond to the trivial representation) and terms which do not  (which have the form $ \cS^m \cT$ and correspond to non trivial characters). Note that products of terms of the form $\cS^m \cT$ may contain a copy of the trivial bundle in their decomposition.

In order to extract the full contribution of the trivial bundle we will formally write 
\be \label{symm-to-rat}
\bigoplus_{k=0}^\infty q^{n k} \cS^k \cT = \frac{1}{1-\cT q^{n} + q^{2 n}} = \exp \left( - \log (1-\cT q^{n} + q^{2 n}) \right) \, .
\ee
More precisely we formally identify $\cT$ with the character of the fundamental representation of SU(2) and interpret the above formula as a formal power series in its generator. 
This rewriting is convenient since now by expanding
we have succeeded in writing $\mathbb{E}$ as a direct sum whose summands are all of the form $\cT^k$ for some $k \in \mathbb{N}$.

Now to extract the contribution of the trivial bundle $\cO$ we have to use repeatedly the tensor product decomposition rules, recalling that $\cT$ transforms as the character of the fundamental representation under the SU(2) action. One starts by $\cT \otimes \cT = \cO \oplus \cS^2 \cT$, where the second factor does not admit global sections and can therefore be discarded. Tensoring again by $\cT$ one gets $\cT^3 = 2 \cT \oplus \cS^3 \cT$. Similarly $\cT^4 = 2 \cO \oplus 3 \cS^2 \cT \oplus \cS^4 \cT$. It is easy to see that any time we tensor  $\cT^k$ with $k$ odd with $\cT$, the resulting decomposition of $\cT^{k+1}$ has one $\cO$ factor whose coefficient is the same as the coefficient of the $\cT$ factor in the decomposition of $\cT^k$. Similarly any time we tensor $\cT^n$ with $n$ even with $\cT$, the decomposition $\cT^{n+1}$ has one $\cT$ factor whose coefficient is the sum of the coefficients of $\cO$ and $S^2 \cT$ in $\cT^n$. The coefficient of $\cS^2 \cT$ in $\cT^n$ is however the sum of the coefficients of $\cT$ and $\cS^3 \cT$ in $\cT^{n-1}$. All these facts follow immediately from the tensor product decomposition of products of the fundamental representation. In summary we see that
\begin{align}
\cT^{2 i} &= C_i \ \cO \oplus \cdots \cr 
\cT^{2 i+ 1} & = C_{i+1} \ \cT \oplus \cdots 
\end{align}
where $i \in \mathbb{N}$ and $C_i = \frac{(2i)!}{(i+1)! i!}$ is the $i^{\rm th}$ Catalan number. The dots denote terms which are sums of factors of the form $\cS^k \cT$ with $k>1$. 
%
%The first few terms are
%\begin{align}
%\cT^2_X &= \cO_X \oplus \cS^2 \cT_X \cr
%\cT^3_X &= 2 \cT_X \oplus \cS^3 \cT_X \cr
%\cT^4_X &= 2 \cO_X \oplus 3 \cS^2 \cT_X \oplus \cS^4 \cT_X \cr
%\cT^5_X &= 5 \cT_X \oplus 4 \cS^3 \cT_X \oplus \cS^5 \cT_X  \cr
%\cT^6_X &= 5 \cO_X \oplus 9 \cS^2 \cT_X \oplus5  \cS^4 \cT_X \oplus \cS^6 \cT_X \cr
%\cT^7_X &= 14 \cT_X \oplus 14 \cS^3 \cT_X \oplus 6 \cS^5 \cT_X  + \cS^7 \cT_X
%\end{align}
%
%We will be able to prove by induction that the coefficient of $\cS^k \cT$ in $\cT^l$, is for $k$ and $l$ even
%\be
%\text{Coeff} \left( \left(\frac{1-\sqrt{1-4 x}}{2 x}\right)^{k+1} x^{k/2} \ ; \  x^{l/2} \right)
%\ee
%while for $k$ and $l$ odd
%\be
%\text{Coeff} \left( \left(\frac{1-\sqrt{1-4 x}}{2 x}\right)^{k+1} x^{(k-1)/2} \ ; \  x^{(l-1)/2} \right)
%\ee
%

We are finally ready to put all of our results together. The Hodge-elliptic genus has two contributions: the contribution from the $\cS^{k >0} \cT_X$ bundles, which is equal to their contribution to the elliptic genus (the $u$ factor cancels with the overall $1/u$ normalization of \eqref{defHEll}) since for these bundles only $H^1$ is non trivial; and the contribution from the $\cO_X$ factors which by Serre duality is equal to their contribution to the elliptic genus weighted by $\frac12 \left( u + \frac{1}{u} \right)$, since for these bundles only $H^0 $ and $H^2$ are non trivial and they are weighted respectively by $1/u$ and $u$.

Therefore an equivalent and perhaps simpler way of computing the Hodge elliptic genus is to compute the ordinary elliptic genus and add the contribution of the trivial bundles weighted by $\frac12 \left( u + \frac{1}{u} \right) -1$. We formally express this by writing
\be
\HEll^c (\tau, z, \nu ) = \mathsf{Ell} (\tau , z ) - \left[ 1- \frac12 \left( u + \frac{1}{u} \right) \right] \mathsf{Ell} (\tau , z ) \bigg\vert_{\cO_X}
\ee
where $ \mathsf{Ell} (\tau , z ) \big\vert_{\cO_X}$ is the contribution to the Elliptic genus from the flat bundle $\cO_X$, and this equation is intended as an equivalence between formal power series.

This strategy can be easily implemented to compute $\HEll^c (\tau, z, \nu)$ as a power series in $q$
\begin{align}
\HEll^c (\tau, z, \nu ) = & \left( u y+\frac{u}{y}+\frac{y}{u}+\frac{1}{u y}+20 \right) \\
& + q \left( u y+\frac{y}{u}+\frac{u}{y}+\frac{1}{u y}-2 u-\frac{2}{u}+20 y^2+\frac{20}{y^2}-130
   y-\frac{130}{y}+220 \right) \cr
   & + q^2 \Big(
   u y^3+\frac{y^3}{u}+\frac{u}{y^3}+\frac{1}{u y^3}-2 u y^2-\frac{2 y^2}{u}-\frac{2
   u}{y^2}-\frac{2}{u y^2}+4 u y+\frac{4 y}{u}+\frac{4 u}{y}+\frac{4}{u y} \cr & -6
   u-\frac{6}{u}+220 y^2+\frac{220}{y^2}-1034 y-\frac{1034}{y}+1628
      \Big) \cr
     & + q^3 \Big(
      u y^3+\frac{y^3}{u}+\frac{u}{y^3}+\frac{1}{u y^3}-6 u y^2-\frac{6 y^2}{u}-\frac{6
   u}{y^2}-\frac{6}{u y^2}+13 u y+\frac{13 y}{u}+\frac{13 u}{y}+\frac{13}{u y} \cr & -16
   u-\frac{16}{u}-130 y^3-\frac{130}{y^3}+1628 y^2+\frac{1628}{y^2}-5530
   y-\frac{5530}{y}+8064
      \Big) + \cdots \nonumber
\end{align}
We will show momentarily how to obtain $\HEll^c (\tau, z, \nu)$ in closed form, by finding a way to implement the above arguments systematically. Before that we have to revisit the computation of the Elliptic genus.

\subsection{The Elliptic genus revisited} \label{Ellrevisited}

%We will now show how the above representation theory arguments can be implemented efficiently in the computation of the Hodge-Elliptic genus. Before doing that, let us reconsider more in detail the Elliptic genus.

The Elliptic genus integrand is determined, up to a normalization, by the Chern character of the bundle \eqref{bundleEll}. By using \eqref{symm-to-rat} we can write
\be \label{integrandEll}
\frac{1}{y} \mathrm{ch} (\mathbb{E} ) = \frac{1}{y} \prod_{n=1}^\infty \frac{
\left(
1 - \frac{y}{\zeta} q^{n-1}
\right)
\left(
1 - \frac{\zeta}{y} q^{n}
\right)
\left(
1 - \frac{1}{y \zeta} q^{n}
\right)
\left(
1 - y \, \zeta q^{n-1}
\right)
}
{
\left(
1 - \zeta q^{n}
\right)^2
\left(
1 - \frac{1}{\zeta} q^{n}
\right)^2
}
\ee
where we have used the splitting principle to write $\mathrm{ch} (\cT) = \zeta + 1/\zeta$, where $\zeta = \e^x$ and $x$ is the integration variable. Now following \cite{Creutzig:2013mqa} we use the  following denominator formulas from \cite{kac} (see Example 4.1)
\begin{align}
\sum_{j \in \zed} \frac{\zeta^j}{1-\frac{q^j}{y}} & = \frac{\zeta}{1-\zeta} \prod_{n=1}^\infty \frac{
\left(
1 - q^n
\right)^2
\left(
1 - \frac{\zeta}{y} q^n
\right)
\left(
1 - \frac{y}{\zeta} q^{n-1}
\right)
}
{
\left(
1 - \zeta \, q^n
\right)
\left(
1 - \frac{1}{\zeta} \, q^n
\right)
\left(
1 - \frac{1}{y} \, q^n
\right)
\left(
1 - y \, q^{n-1}
\right)
}
\cr
\sum_{j \in \zed} \frac{\zeta^j}{1-y \, q^j} & = \frac{1}{1-\zeta} \prod_{n=1}^\infty \frac{
\left(
1 - q^n
\right)^2
\left(
1 - \zeta \, y \, q^{n-1}
\right)
\left(
1 - \frac{1}{y\,\zeta} q^{n}
\right)
}
{
\left(
1 - \zeta \, q^n
\right)
\left(
1 - \frac{1}{\zeta} \, q^n
\right)
\left(
1 - \frac{1}{y} \, q^n
\right)
\left(
1 - y \, q^{n-1}
\right)
}
\end{align}
Noting that
\begin{align}
\theta_1^2 (\tau , z) & = - q^\frac14 \frac{1}{y} \prod_{n=1}^\infty \left( 1 - q^n \right)^2 \left( 1 - y \, q^{n-1} \right)^2 \left( 1 - \frac{1}{y} q^n \right)^2 
\cr
\eta^6 (\tau) & = q^\frac14 \prod_{n=1}^\infty \left( 1 - q^n \right)^6
\end{align}
we can now write \eqref{integrandEll} as
\begin{align} \label{zetaNsN}
\frac{1}{y} \mathrm{ch} (\mathbb{E}) & = -\frac{(1-\zeta)^2}{\zeta} \  \frac{\theta_1^2 (\tau,z)}{\eta(\tau)^6} \ \sum_{i,j \in \zed} \frac{\zeta^{i+j}}{(1-y \, q^j) (1 - \frac{1}{y} q^i)}
 = - \frac{\theta_1 (\tau,z)^2}{\eta (\tau)^6} \ \sum_{N \in \zed} \sum_{i \in \zed} \frac{\zeta^{N-1}- 2 \zeta^N+ \zeta^{N+1}}{(1-y \, q^i) (1 - \frac{1}{y} q^{N-i})}  
\nonumber  \\ & 
 = -  \frac{\theta_1 (\tau,z)}{\eta (\tau)^3} \sum_{N \in \zed} \zeta^N (s_{N+2}(\tau , z) - 2 s_{N+1}(\tau , z) + s_{N}(\tau , z)) 
\end{align}
by changing summation variable $i+j=N$. %Note that $- \theta_1^2 / \eta^6 = \phi_{-2,1}$
In the last step we have introduced the functions
\be
s_N (\tau , z) = \frac{\theta_1 (\tau , z)}{\eta (\tau)^3} \sum_{i \in \zed} \frac{1}{(1-y q^i)(1-\frac{1}{y} q^{N-1-i})} \, .
\ee
%we can write
%\be
%- y  \frac{\theta_1^2 (q,y)}{\eta^6 (q)} \ \sum_{N \in \zed} \sum_{i \in \zed} \frac{z^{N-1}- 2 z^N+ z^{N+1}}{(1-y \, q^i) (1 - \frac{1}{y} q^{N-i})}  
% = -y  \frac{\theta_1 (q,y)}{\eta^3 (q)} \sum_{N \in \zed} z^N (f_{N+1} - 2 f_N + f_{N-1})
%\ee
As we have discussed we would like to rewrite the Chern character of $\mathbb{E}$ in a form where the contribution from each $S^N \cT$ bundle is highlighted:
\be \label{chSNFN}
\frac{1}{y} \ch (\mathbb{E}) = \sum_{N=0}^\infty \ch (S^N \cT) \mathscr{F}_N (\tau , z)
\ee
in terms of certain functions $\mathscr{F}_N$ to be determined by comparing with \eqref{chSNFN}. In order to do so note that on the right hand side of \eqref{chSNFN} the coefficient of $\zeta^k$ is of the form $\sum_{i=k}^\infty \mathscr{F}_{2 i}$. Therefore it follows that we can obtain $\mathscr{F}_N$ by taking the difference between the coefficient of $\zeta^N$ and the coefficient of $\zeta^{N+2}$ in \eqref{zetaNsN}, since all the other terms cancel. Therefore
\be \label{functionF}
 \mathscr{F}_N (\tau , z)= - \frac{\theta_1 (\tau , z)}{\eta (\tau)^3} \left( s_{N} (\tau , z) - 2 s_{N+1}(\tau , z) + 2 s_{N+3} (\tau , z)- s_{N+4} (\tau , z)\right)
\ee
In particular now we know how to isolate the trivial character which corresponds to the contribution of the trivial bundle.

We prove in Appendix \ref{sidentities} that
\be \label{functions}
s_N (\tau , z) = \frac{\theta_1 (\tau, z)}{\eta(\tau)^3} \widetilde{s}_N (\tau) + \delta_{N,1} \, \theta_1 (\tau,z) \, \mu (\tau,z)
\ee
with
%\be \label{functionstilde}
%\widetilde{s}_N (\tau) = \left\{
%\begin{matrix}
%\frac{q}{1-q} & \text{if} \ N=0 \\
%2 F_2^{(2)} (\tau) & \text{if} \ N=1 \\
%\frac{N-1}{1-q^{N-1}} & \text{otherwise} 
%\end{matrix}
%\right.
%\ee
\be \label{functionstilde}
\widetilde{s}_N (\tau) = 
\begin{dcases}
\frac{q}{1-q} & \text{if } \ N=0 \, ,\\
2 F_2^{(2)} (\tau) & \text{if } \ N=1 \, ,\\
\frac{N-1}{1-q^{N-1}} & \text{otherwise.} 
\end{dcases}
\ee
Therefore we can write the full Elliptic genus as
\be \label{EllFrep}
\mathsf{Ell} (\tau , z) = \frac{1}{y} \int \ch (\mathbb{E}) \mathrm{Todd} (\cT) = \sum_{N=0}^\infty \mathscr{F}_N (\tau , z) \int \ch (S^N \cT) \mathrm{Todd} (\cT) \, .
\ee
The integration is now elementary:
\begin{align}
\int \ch (S^N \cT) \mathrm{Todd} (\cT) & = \int \frac{\e^{(N+1)x} - \e^{- (N+1) x}}{\e^x - \e^{-x}} \frac{-x^2}{(1-\e^x)(1-\e^{-x})} 
\cr &
= \frac{1}{12}  (2 N^3 + 6 N^2 + 3 N -1) \int x^2 = -2  (2 N^3 + 6 N^2 + 3 N -1) 
\end{align}
where we have used \eqref{intx24}.

We conclude that the Elliptic genus can be written as
\be
\mathsf{Ell} (\tau , z) =- \sum_{N=0}^\infty 2  (2 N^3 + 6 N^2 + 3 N -1) \ \mathscr{F}_N (\tau , z) \, .
\ee
It is convenient to rearrange terms as
\begin{align}
\sum_{N=0}^\infty 2  (2 N^3 + 6 N^2 + 3 N -1) \ (s_N (\tau , z) - 2 s_{N+1} (\tau , z) + 2 s_{N+3} (\tau , z) - s_{N+4} (\tau , z) ) 
\cr
= - 2 s_0 (\tau , z) + 24 s_1 (\tau , z)  + 50 s_2 (\tau , z)  + 48 \sum_{N=3}^\infty s_N (\tau , z)  \, .
\end{align}
The first three contributions can be checked directly. All the other follow from some boring but straightforward algebra
\begin{align}
 \left( 2  (2 N^3 + 6 N^2 + 3 N -1) \right) - 2 \left( 2  (2 (N-1)^3 + 6 (N-1)^2 + 3 (N-1) -1) \right) \\ \nonumber
 + 2 \left( 2  (2 (N-3)^3 + 6 (N-3)^2 + 3 (N-3) -1) \right) \\
 - \left( 2  (2 (N-4)^3 + 6 (N-4)^2 + 3 (N-4) -1) \right) = 48 \, . \nonumber
\end{align}
To summarize we have shown
\be
\mathsf{Ell} (\tau , z) =\frac{\theta_1 (\tau , z)^2}{\eta (\tau)^3} \left[
\frac{1}{\eta (\tau)^3} \left(
- 2 \widetilde{s}_0 + 24 \widetilde{s}_1 + 50 \widetilde{s}_2 + 48 \sum_{N=3}^{\infty} \widetilde{s}_N 
\right) + 24 \mu (\tau , z)
\right] \, .
\ee
Furthermore the combination
\begin{align}
- 2 \widetilde{s}_0 + 50 \widetilde{s}_2 + 48 \sum_{N=3}^{\infty} \widetilde{s}_N = 2 + 48 \frac{1}{1-q} + 48 \sum_{n=2}^\infty \frac{n}{1-q^n} = - 2 E_{2} (\tau) \, ,
\end{align}
gives the Eisenstein series $E_2 (\tau)$. Indeed the latter can be written as\footnote{Note that this identity does \textit{not} hold as a truncated expansion, because of the analytical continuation.}
\begin{align} \label{E2ancont}
E_2 &= 1 - 24 \sum_{n=1}^\infty \frac{n q^n}{1-q^n} = 1 - 24 \sum_{n=1}^\infty \left( \frac{n}{1-q^n} - n \frac{1-q^n}{1-q^n} \right) = 
1 - 24 \left( \sum_{n=1}^\infty \frac{n}{1-q^n} + \frac{1}{12} \right) \cr
& = -1 -24 \sum_{n=1}^\infty \frac{n}{1-q^n}
\end{align}
where we have used $\sum_{n=1}^{\infty} n = \zeta (-1) = - \frac{1}{12}$ in terms of the analytically continued Riemann zeta function $\zeta (s)$.
%We conclude that\footnote{Note that this identity does \textit{not} hold as a truncated expansion, because of the above analytical continuation.}
%\begin{align}
%- 2 \widetilde{s}_0 + 50 \widetilde{s}_2 + 48 \ \sum_{N=3}^{\infty} \widetilde{s}_N = - 2 E_{2} (\tau)
%\end{align}
By putting everything together we finally have
\be
\mathsf{Ell} (\tau , z) = \frac{\theta_1 (\tau , z)^2}{\eta (\tau)^3} \left( \frac{- 2 E_2 (\tau) + 48 F_2^{(2)} (\tau)}{\eta (\tau)^3} + 24 \mu (\tau , z) \right)
\ee
as expected.

\subsection{The complex Hodge-Elliptic genus} \label{complexHEll}

We have just shown that
\begin{align}
\mathsf{Ell} (\tau , z) &= \frac{1}{y} \chi (\mathbb{E}) = \frac{1}{y} \sum_{j=0}^2 (-1)^j \dim H^j (\mathbb{E}) 
= \sum_{N=0}^{\infty} \mathscr{F}_N (\tau , z) \sum_{j=0}^2 (-1)^j \dim H^j (S^N \cT) \, .
\end{align}
Similarly we can write for the complex Hodge-Elliptic genus
\begin{align} \label{HEll-Ell-Euler}
\HEll^c (\tau , z , \nu) &=  \frac{1}{y u} \sum_{j=0}^2 (-u)^j \dim H^j (\mathbb{E}) 
\cr & = \sum_{N=0}^{\infty} \mathscr{F}_N  (\tau , z)  \left( \frac{1}{u} \dim H^0 \left( S^N \cT \right) -  \dim H^1 \left( S^N \cT \right) + u  \dim H^2 \left( S^N \cT \right) \right)
\cr & = \frac12 \left( u + \frac{1}{u} \right) \mathscr{F}_0  (\tau , z) \, \chi (\cO) + \sum_{N=1}^\infty \mathscr{F}_N  (\tau , z)  \, \chi (S^N \cT)
\cr & = \mathsf{Ell} (\tau , z)  - \left[ 1 - \frac12 \left( u + \frac{1}{u} \right) \right] \mathscr{F}_0  (\tau , z) \, \chi (\cO) 
\end{align}
where we have used
\be
\frac{1}{u} \dim H^0 \left( S^N \cT \right) -  \dim H^1 \left( S^N \cT \right) + u  \dim H^2 \left( S^N \cT \right) = \left\{
\begin{matrix}
\frac12 \left( \frac{1}{u} + u \right) \chi (\cO) & \text{if $N=0$} \\ & \\
\chi (S^N \cT) & \text{if $N \neq 0$}
\end{matrix}
\right.
\ee
Now we can compute the trivial bundle contribution
\begin{align}
& \left[ 1 - \frac12 \left( u + \frac{1}{u} \right) \right] \mathscr{F}_0  (\tau , z) \, \chi (\cO) \\
& = \left[ 1 - \frac12 \left( u + \frac{1}{u} \right) \right] \left[
\left( - 2 \frac{\theta_1 (\tau , z)^2}{\eta (\tau)^6} \right) \left(
\frac{q}{1-q} + \frac{4}{1-q^2} - \frac{3}{1-q^3} - 4 F_2^{(2)} (\tau)
\right)
+ 4 \frac{\theta_1 (\tau , z)^2}{\eta (\tau)^3} \, \mu (\tau , z)
\right]
\cr \nonumber
& = \left[ -2 + u + \frac{1}{u} \right] \frac{\theta_1 (\tau , z)^2}{\eta (\tau)^6} \left(
\frac{1+2 q+6 q^2+2 q^3+q^4}{(1-q)(1+q) (1+q+q^2)} - 4 F_2^{(2)} (\tau)
\right) 
\cr \nonumber & \qquad 
+  \left[ 4 - 2 \left( u + \frac{1}{u} \right) \right] \frac{\theta_1 (\tau , z)^2}{\eta (\tau)^3} \, \mu(\tau , z)
\end{align}
Finally putting everything together we conclude that 
\begin{align} \label{HEll1}
\HEll^c (\tau, z, \nu ; K3) &=   \frac{\theta_1 (q,y)^2}{\eta (q)^3}  \left[ \left( 20 + 2 \left( \frac{1}{u} + u \right) \right) \mu (q,y) +   H(q) \right] \cr & +  \left( 2 -  \left( \frac{1}{u} + u \right) \right)  \frac{\theta_1 (q,y)^2}{\eta (q)^6} \left( \frac{1+2 q+6 q^2+2 q^3+q^4}{(1-q)(1+q) (1+q+q^2)} - 4 F_{2}^{(2)} (\tau)\right)
\end{align}
Equivalently we can use the definition \eqref{FunctionH} to write
%\be
%\frac{4 F_2^{(2)} (\tau)}{\eta (\tau)^3} = \frac{1}{12} \left( H (\tau) + \frac{2 E_2 (\tau)}{\eta (\tau)^3} \right)
%\ee
%we can also write
\begin{align} \label{HEll2}
\HEll^c (\tau, z, \nu ; K3) &=   \frac{1}{24} \left( 20 + 2 \left( u + \frac{1}{u} \right) \right) \mathsf{Ell} (\tau , z) \cr & +  \left( 2 -  \left( \frac{1}{u} + u \right) \right)  \frac{\theta_1 (\tau,z)^2}{\eta (\tau)^6} \left(\frac{1+2 q+6 q^2+2 q^3+q^4}{(1-q)(1+q) (1+q+q^2)}- \frac16 E_2 (\tau) \right)
\end{align}
Recall that $\mu (\tau ,z)$ and $H (\tau)$ are mock modular forms with shadows $- \eta (\tau)^3$ and $24 \eta (\tau)^3$. Remarkably the $u$-dependence factors out in the first line and again the two shadows cancel exactly to give a Jacobi form with a $u$-dependent coefficient which reduces to one in the limit $u \rightarrow 1$. In the second line $\varphi_{-2,1} = - \theta_1^2 / \eta^6$ is also a Jacobi form, and $E_2 (\tau)$ is a quasi-modular form, it transforms as a modular form of weight 2 when we add $-3/ \pi \mathrm{Im} (\tau)$. Modular properties are however spoiled by the presence of the rational function.

Note that the form \eqref{HEll2} of the $u$-dependent factor $ \frac{1}{24} \left( 20 + 2 u + \frac{2}{u} \right) $ multiplying the Elliptic genus plus a correction is common to all Hodge-Elliptic genera.

%Note that the rational function can also be written as
%\be
%\frac{q^4 + 2 q^3 + 6 q^2 + 2 q + 1}{(q-1) (q+1) (q^2+ q + 1)} = (q^4 + 2 q^3 + 6 q^2 + 2 q + 1) 
%\sum_{n=0}^{\infty} \left( -q^{6n} +q^{6n+1} - q^{6n+2} \right) = \sum_{n=0}^\infty \alpha_n q^n
%\ee
%where the coefficients $\alpha_n$ can be determined explicitly.

\subsection{Character decomposition and Mathieu Moonshine}

%WE POINT OUT THAT HELL CAN BE WRITTEN AS A REFINEMENT OF THE ELL DECOMPOSITION, different from benjamin D1/D5. Explain why, as a curiosity?

Among the connections between string theory and number theory, perhaps one of the most striking is the observation by Eguchi, Ooguri and Tachikawa   \cite{Eguchi:2010ej} that certain coefficients in the character expansion of the Elliptic genus are twice the dimensions of certain irreducible representations of the largest Mathieu group $\mathbb{M}_{24}$. We would like to investigate how this statement is modified when the Elliptic genus is refined into its Hodge-Elliptic counterpart. The main motivation in doing so is that often in string theory refined enumerative invariant help to understand the Hilbert space of BPS states\footnote{On general ground one expects that the categorification of Donaldson-Thomas invariants will provide a model for the Hilbert space of BPS states. Making sense of this statement is one of the goals of Donaldson-Thomas theory.} and could help in identifying the physical and geometrical reason the Mathieu group appears. We will now show directly that all the Hodge-Elliptic genera we have seen can be decomposed as a sum over the same superconformal characters as the Elliptic genus with $u$ dependent coefficients.

The Elliptic genus admits the following decomposition \cite{Eguchi:2010ej} 
\be \label{Ellcharacters}
\mathsf{Ell} (\tau ; z) = 20 \, \ch_{\frac14,0}  (\tau ; z) - 2 \, \ch_{\frac14,\frac12} (\tau ; z) + \sum_{n=1}^{\infty} \, c_H (n) \, \ch_{\frac14 + n , \frac12} (\tau , z)
\ee
in terms of the characters of the superconformal algebra. These have the form
\be
\mathrm{ch}_{h,\ell} (\tau , z) = \mathrm{Tr}_{V_{h,\ell}} \left( 
(-1)^{J_0} \, y^{J_0} q^{L_0 - \frac{c}{24}}
\right) \, ,
\ee
where an irreducible representation $V_{h , \ell}$ of the $\cN=4$ algebra is labelled by the quantum numbers $h$ and $\ell$, the eigenvalues of $L_0$ and $J_0^3$ respectively. For central charge $c=6$ the massless representations have quantum numbers $(h,\ell) = (\frac14 , 0)$ and $(h,\ell) = (\frac14 , \frac12)$, while the massive representations have $(h,\ell) = (\frac14 + n , 0)$ with $n=1,2,\dots$. Their characters are \cite{Eguchi:1987sm,Eguchi:1987wf,Eguchi:1988af}
\begin{align}
\ch_{\frac14,0}  (\tau ; z) & = \frac{\theta_1 (\tau ; z)^2}{\eta(\tau)^3} \ \mu (\tau , z) \, ,  \cr
\ch_{\frac14,\frac12} (\tau ; z) & = q^{-\frac18} \frac{\theta_1 (\tau ; z)^2}{\eta(\tau)^3} - 2 \frac{\theta_1 (\tau ; z)^2}{\eta(\tau)^3} \ \mu(\tau , z) \, , \cr
\ch_{\frac14 + n , \frac12} (\tau ; z) & = q^{-\frac18 + n} \frac{\theta_1 (\tau ; z)^2}{\eta(\tau)^3} \, .
\end{align}
In \eqref{Ellcharacters} the coefficients $c_H (n)$ are defined via \eqref{FunctionH}, or equivalently
%\be
%H (\tau) = q^{-\frac18} \left( -2 + \sum_{n=1}^\infty c_H (n) q^n \right) = 2 q^{-\frac18} \left( -1 + 45 q + 231 q^2 + 770 q^3 + \cdots \right) 
%\ee
%or equivalently
\be \label{rewriteH}
\sum_{n=1}^{\infty} \, c_H (n) \, \ch_{\frac14 + n , \frac12} (\tau , z) =  \frac{\theta_1 (\tau ; z)^2}{\eta(\tau)^3} H (\tau) + 2  q^{-\frac18}  \frac{\theta_1 (\tau ; z)^2}{\eta(\tau)^3} \, .
\ee

Now let us consider the Hodge-Elliptic genera \eqref{HEll2}, \eqref{HEllorb} and \eqref{genericHEll}. Differently from \eqref{Ellcharacters} the Hodge-Elliptic genus will involve also right-moving characters, of the form 
\be
\mathrm{ch}_{h,\ell} (\overline{\tau} , \nu) = \mathrm{Tr}_{V_{h,\ell}} \left( 
(-1)^{\overline{J}_0} \, u^{\overline{J}_0} \overline{q}^{\overline{L}_0 - \frac{\overline{c}}{24}}
\right) \, .
\ee
However since the sum is constrained to states with $\overline{L}_0 = \frac{\overline{c}}{24}$, the net effect is that we expect a decomposition similar to that of \eqref{Ellcharacters} whose coefficients depend now on the fugacity $u$. In the following we will exhibit explicitly this structure for the three Hodge-Elliptic partition function we know in closed form. The results have the form of a refined Mathieu moonshine. It is natural to hope that the $u$-dependent coefficients can now be interpreted as a refined trace over the $\mathbb{M}_{24}$ module which appear in the decomposition \eqref{Ellcharacters}. Indeed from the geometrical point of view in the complex and generic definitions \eqref{defHEll} and \eqref{genericHEll2} the refinement consists in an extra parameter which weight differently the cohomology groups. We will see that this structure is preserved in the holomorphic character decomposition.

Consider first the complex Hodge-Elliptic genus \eqref{HEll2}. Using the definitions of the characters and \eqref{rewriteH} we see that
\begin{align}
\HEll^c (\tau , z , \nu) & = 20 \, \ch_{\frac14,0} (\tau ; z) + 2 \left( \frac{1}{u} + u \right) \frac{\theta_1 (\tau,z)^2}{\eta (\tau)^3}  \mu (\tau,z) + 
\cr & + \frac{1}{24} \left( 20 + 2 \left(u + \frac{1}{u} \right) \right) \left[  \sum_{n=1}^{\infty} \, c_H (n) \, \ch_{\frac14 + n , \frac12} (\tau , z)  -  2  q^{-\frac18}  \frac{\theta_1 (\tau ; z)^2}{\eta(\tau)^3} \right]
 \cr & +
\left( 2 -  \left( \frac{1}{u} + u \right) \right)  \frac{\theta_1 (\tau,z)^2}{\eta (q)^6} \left( \frac{1+2 q+6 q^2+2 q^3+q^4}{(1-q)(1+q) (1+q+q^2)}- \frac16 E_2 (\tau) \right) \, .
\end{align}
Adding and subtracting $(\frac{1}{u} + u) q^{- \frac18}  \frac{\theta_1 (\tau,z)^2}{\eta (\tau)^3}$ we can write
\begin{align}
\HEll^c (\tau , z , \nu) &=20\,  \ch_{\frac14,0} (\tau ; z) - \left( \frac{1}{u} + u \right) \ch_{\frac14,\frac12} (\tau ; z) 
\\ & + \frac{1}{24} \left( 20 + 2 u + \frac{2}{u}  \right) \sum_{n=1}^{\infty} \, c_H (n) \, \ch_{\frac14 + n , \frac12}  (\tau , z) \cr & 
+  \left( 2 - \frac{1}{u} - u  \right) \, q^{-\frac18}  \frac{\theta_1 (\tau , z)^2}{\eta (\tau)^3} \left[ \frac{q^{1/8}}{\eta(\tau)^3} \left( \frac{1+2 q+6 q^2+2 q^3+q^4}{(1-q)(1+q) (1+q+q^2)} - \frac16 \, E_2 (\tau) \right) - \frac56 \right] \nonumber
\end{align}
We can interpret the second line as a correction to the coefficients $c_H (n)$.
%\be
%\left( 2 -  \left( \frac{1}{u} + u \right) \right) q^{-\frac18}  \frac{\theta_1 (q,y)^2}{\eta (q)^3} \left[ \frac{q^{\frac18}}{\eta(q)^3} \left( \widetilde{h}_0 (q) - 2 \widetilde{h}_1 (q) + 2 \widetilde{h}_3 (q) -\widetilde{h}_4 (q) \right) - 1 \right]
%\ee
Indeed it is clear that the terms in square brackets has a $q$ expansion without constant term, since in the $q$ expansion the zeroth order terms coming from the rational function and from the Eisenstein series precisely cancel the $5/6$ factor:
\begin{align}
&  \left[ \frac{q^{1/8}}{\eta(\tau)^3} \left( \frac{1+2 q+6 q^2+2 q^3+q^4}{(1-q)(1+q) (1+q+q^2)}- \frac16 \, E_2 (\tau) \right) - \frac56 \right] = \sum_{n=1}^{\infty} a_H (n) q^n  \cr
 & = \frac{15}{2} q +\frac{79}{2} q^2 +\frac{385}{3} q^3 +\frac{761}{2} q^4 +969 q^5+\frac{13927}{6} q^6+\frac{10293}{2} q^7+10936 q^8+ \cdots  \, .
\end{align}
Therefore the whole expression can be understood as a correction to the coefficients of the characters $\ch_{\frac14 + n , \frac12}$. We can write the complex Hodge-Elliptic genus as
\begin{align}
\HEll^c (\tau , z , \nu) &= 20 \, \ch_{\frac14,0} (\tau : z) - \left( \frac{1}{u} + u \right) \ch_{\frac14,\frac12} (\tau ; z) +  \sum_{n=1}^{\infty} \,\tilde{c}_H (n) \, \ch_{\frac14 + n , \frac12} (\tau , z)
\end{align}
where now the coefficients
\be \label{ctilde1}
\tilde{c}_H (u ; n) = \left(   \frac{1}{24} \left( 20 + 2 u + \frac{2}{u}  \right) c_H (n) + \left( 2 - \frac{1}{u} - u \right) \, a_H (n) \right)
\ee
are functions of $u$.

Equivalently we could have started from the expression \eqref{HEll1} and obtained
\begin{align}
\HEll^c (\tau , z , \nu) &=20 \, \ch_{\frac14,0} (\tau ; z) - \left( \frac{1}{u} + u \right) \ch_{\frac14,\frac12} (\tau ; z) + \sum_{n=1}^{\infty} \, c_H (n) \, \ch_{\frac14 + n , \frac12} (\tau , z)  \\ \nonumber & \!\!\!\! \!\!\!\!
+  \left( 2 -  \frac{1}{u} - u  \right) \, q^{-\frac18}  \frac{\theta_1 (\tau , z)^2}{\eta (\tau)^3} \left[ \frac{q^{1/8}}{\eta(\tau)^3} \left( \frac{1+2 q+6 q^2+2 q^3+q^4}{(1-q)(1+q) (1+q+q^2)}- 4 \, F_2^{(2)} (\tau) \right) - 1 \right] 
\end{align}
Now the term in square brackets has an \textit{integral} expansion in $q$ without constant term
\begin{align}
&  \left[ \frac{q^{1/8}}{\eta(\tau)^3} \left( \frac{1+2 q+6 q^2+2 q^3+q^4}{(1-q)(1+q) (1+q+q^2)} - 4 \, F_2^{(2)} (\tau) \right) - 1 \right]  = \sum_{n=1}^{\infty} b_H (n) q^n  \cr
 & = q^2+q^4+3 q^5+2 q^6+6 q^7+11 q^8+13 q^9+24 q^{10}+ \cdots  
\end{align}
so that the whole expression can be understood as a shift of the coefficients of the characters $\ch_{\frac14 + n , \frac12}$. We can therefore write the Hodge-Elliptic genus as
\begin{align}
\HEll &= 20 \, \ch_{\frac14,0} (\tau : z) - \left( \frac{1}{u} + u \right) \ch_{\frac14,\frac12} (\tau ; z) +  \sum_{n=1}^{\infty} \,\tilde{c}_H (n)  \, \ch_{\frac14 + n , \frac12} (\tau , z)
\end{align}
where now
\be
\tilde{c}_H (u ; n) =  \left(  c_H (n) + \left( 2 - \frac{1}{u} - u \right) \, b_H (n) \right)
\ee
and clearly coincide with \eqref{ctilde1}. Regardless of which expression for these coefficients we find more convenient, we can write down the first few terms
\begin{align}
& \sum_{n=1}^{\infty} \,\tilde{c}_H (u ; n) \, \ch_{\frac14 + n , \frac12} = 2 \Big[
45 q+q^2 \left(232-\frac{u}{2}-\frac{1}{2 u}\right)+770 q^3+q^4
   \left(2278-\frac{u}{2}-\frac{1}{2 u}\right)
   \\ \nonumber &
   +q^5 \left(5799-\frac{3 u}{2}-\frac{3}{2
   u}\right)+q^6 \left(13917-u-\frac{1}{u}\right)+q^7 \left(30849-3
   u-\frac{3}{u}\right)
   \\ \nonumber & 
   +q^8 \left(65561-\frac{11 u}{2}-\frac{11}{2 u}\right)+q^9
   \left(132838-\frac{13 u}{2}-\frac{13}{2 u}\right)+q^{10} \left(260592-12
   u-\frac{12}{u}\right) + \cdots \Big]  \, .
\end{align}
%
%
%\begin{align}
%& \sum_{n=1}^{\infty} \,\tilde{c}_H (u ; n) \, \ch_{\frac14 + n , \frac12} 
%= 90 q+q^2 \left(-u-\frac{1}{u}+464\right)+1540 q^3 \cr
%& +q^4
%   \left(-u-\frac{1}{u}+4556\right)+q^5 \left(-3 u-\frac{3}{u}+11598\right)+q^6 \left(-2
%   u-\frac{2}{u}+27834\right)+
%   \cr &
%   q^7 \left(-6 u-\frac{6}{u}+61698\right)+q^8 \left(-11
%   u-\frac{11}{u}+131122\right)+q^9 \left(-13 u-\frac{13}{u}+265676\right) 
%   \cr &
%   + q^{10}
%   \left(-24 u-\frac{24}{u}+521184\right)+O\left(q^{11}\right)
%\end{align}
It remains to be understood if these coefficients have any interpretation in the context of Mathieu Moonshine.

Also the orbifold Hodge-Elliptic genus \eqref{HEllorb} can be written in terms of the superconformal characters
\begin{align}
\HEll^{\mathrm{orb}} (\tau , z, \nu) = & \frac{1}{24} \left( \frac{2}{u} + 20 + 2 u \right) \left[ 20 \, \ch_{\frac14,0} (\tau , z) - 2 \ch_{\frac14 , \frac12} (\tau , z) + \sum_{n=1}^\infty c_H (n) \ch_{\frac14+n,\frac12} (\tau , z) \right] 
\nonumber \\ & 
+ \left( 1 - \frac{1}{2 u} - \frac{u}{2} \right) \frac{\theta_1 (\tau , z)^2}{\eta (\tau)^3} q^{-\frac18} \left[
\frac{q^{\frac18}}{\eta (\tau)^3} \left(
1 +  8 \Lambda_2 (\tau)
\right)
\right] \, . 
\end{align}
The term on the second line has a $q$ expansion which starts with a constant term. We can however write it in terms of the superconformal characters as
\begin{align}
\frac{\theta_1 (\tau , z)^2}{\eta (\tau)^3} q^{-\frac18} \left[
\frac{q^{\frac18}}{\eta (\tau)^3} \left(
1 +  8 \Lambda_2 (\tau)
\right)
\right] = \frac{\theta_1 (\tau , z)^2}{\eta (\tau)^3} q^{-\frac18} \left[
\frac53 +  \sum_{n=1}^\infty d_H (n) q^n
\right] 
\cr = \frac53 \left( \ch_{\frac14 , \frac12} (\tau , z) + 2 \ch_{\frac14 , 0} (\tau , z) \right)  + \sum_{n=1}^\infty d_H (n) \ch_{\frac14 + n , \frac12} (\tau , z)
\end{align}
where
\be
\sum_{n=1}^\infty d_H (n) q^n = 21 q+79 q^2+\frac{878}{3} q^3+789 q^4+2068 q^5+\frac{14449}{3} q^6 + \dots %+10779 q^7+22566 +  q^8+45873 q^9+89328 q^{10}+ \dots
\ee
which has no constant term. Putting everything together the orbifold Hodge-Elliptic genus has the simple expression
\begin{align}
\HEll^{\mathrm{orb}} (\tau , z, \nu) &= 20 \, \ch_{\frac14,0} (\tau , z) - \left( \frac{1}{u} + u \right) \ch_{\frac14,\frac12} (\tau , z) +  \sum_{n=1}^{\infty} \,\tilde{c}^{\mathrm{orb}}_H (n)  \, \ch_{\frac14 + n , \frac12} (\tau , z)
\end{align}
where
\begin{align}
\tilde{c}_H (u , n) & =  \frac{1}{24} \left( \frac{2}{u} + 20  + 2 u \right) c_H (n) + \left( 1 - \frac{1}{2 u} - \frac{u}{2} \right) d_H (n) \cr &
= 2 \Big[
q \left(48-\frac{3 u}{2}-\frac{3}{2 u}\right)+q^2 \left(232-\frac{u}{2}-\frac{1}{2
   u}\right)+q^3 \left(788-9 u-\frac{9}{u}\right)
   \cr & \ \
   +q^4 \left(2292-\frac{15
   u}{2}-\frac{15}{2 u}\right)+q^5 \left(5864-34 u-\frac{34}{u}\right)+q^6
   \left(14004-\frac{89 u}{2}-\frac{89}{2 u}\right) 
   \cr & \ \ 
   +q^7 \left(31092-\frac{249
   u}{2}-\frac{249}{2 u}\right)+q^8 \left(65908-179 u-\frac{179}{u}\right)
   \cr & \ \
   +q^9
   \left(133624-\frac{799 u}{2}-\frac{799}{2 u}\right)
   +q^{10} \left(261804-618
   u-\frac{618}{u}\right)+\cdots 
     \Big]
\end{align}

Finally let us consider the generic Hodge-Elliptic genus \eqref{genericHEll}
\begin{align}
\HEll^g (\tau , z , \nu) & =  \mathsf{Ell} (\tau , z ) + \left( 2 - \frac{1}{u} - u \right) \mathrm{ch}_{\frac14 , \frac12} (\tau , z)
\cr & =   20 \, \ch_{\frac14,0}  (\tau ; z) - \left(u + \frac{1}{u} \right) \ch_{\frac14,\frac12} (\tau ; z) + \sum_{n=1}^{\infty} \, c_H (n) \, \ch_{\frac14 + n , \frac12} (\tau , z) \, .
\end{align}
In the case only the coefficient of $ \ch_{\frac14,\frac12} (\tau ; z) $ is modified.

Overall we find that the Hodge-Elliptic genera we have consider always admit a character expansion of the form \eqref{Ellcharacters} but where now the coefficients are $u$ dependent.

\section{Refined Dyons and enumerative geometry} \label{Enumerative}

In this Section we will discuss a refined version of the counting function $\Phi_{10}$ and use it to make some predictions for enumerative geometry. We will discuss this procedure in general and then we will specialize to the several Hodge-Elliptic genera we have encountered in the previous Section. To each one we can associate a different set of enumerative invariants, conjecturally related by wall-crossing. 

\subsection{A refined BPS counting function}

Now that we have an explicit form for the Hodge-Elliptic partition function, we can use it to define a refined version of the Igusa cusp form $\Phi_{10}$. We expand the Hodge-Elliptic genus as
\be
\HEll (\tau, z, \nu) = \sum_{s,t,v} c (s , t , v) \, q^s \, y^t \, u^v \, ,
\ee
and use the coefficients of the expansion to define the function
\begin{align}
\Phi^{\mathrm{ref}} (\tau , z , \nu, \sigma) & = p\, q \, y \prod_{(s,t,r,v) > 0} \left( 1 - q^s y^t p^r u^v \right)^{c (r\, s ,  t  , v)}
 \cr  & = p\, q \, y \left( 1-  \frac{u}{y} \right) \left( 1 - \frac{1}{u y} \right)  \prod_{\stackrel{s > 0 , r > 0}{(t,v)\in \zed}} \left( 1 - q^s y^t p^r u^v \right)^{c (r\, s ,  t  , v)} \, ,
\end{align}
where the notation $(s,t,r,v) > 0$ means that the product is over all the $t, v \in \zed$ and $s,r \ge 0$ such that one of the following two conditions hold
\begin{itemize}
\item $s > 0$ or $r >0$,
\item $s=r=0$ and $t < 0$.
\end{itemize}
Note that by definition the range of $v$ is just $v = -1,0,1$ since otherwise the coefficients $c (r\, s ,  t  , v)$ vanish, by direct inspection of the Hodge-Elliptic genera. Of course the function $\Phi^{\mathrm{ref}} $ will contain arbitrary powers of $u$. We also stress that the function $\Phi^{\mathrm{ref}} $ so defined, including the range of indices, is again formally symmetric for the exchange of $q$ and $p$. 

Then the main conjecture of \cite{Kachru:2016igs} is that 
\be \label{PhirefDTref}
\Phi^{\mathrm{ref}} (\tau , z , \nu, \sigma) = - \frac{1}{\DT^{\mathrm{ref}}} \, 
\ee
where
\be
\DT^{\mathrm{ref}} = \sum_{h=0}^\infty \DT^{\mathrm{ref}}_h (\mathrm{K3} \times E) \, p^{h-1} = \sum \DT^{\mathrm{ref}}_{h,d,n} \ p^{h-1} q^{d-1} (-y)^{n}
\ee
is the generating function of (reduced) refined Donaldson-Thomas invariants for $K3 \times E$.

This conjecture requires however some explanation. Since the Hodge-Elliptic genus depends on the K\"ahler moduli, the same is true for the generating function of refined BPS invariants. While the unrefined partition function already has some mild wall-crossing behaviour when crossing codimension 1 walls of marginal stability, the refined partition function is expected to be sensitive also to BPS jumping loci, higher codimension loci in the moduli space where certain extra BPS states appear, leaving the rest of the spectrum unchanged \cite{Kachru:2016igs,Kachru:2017yda,Kachru:2017zur}. Across such loci the full generating function is expected to change also for $\cN=4$ compactifications.

The origin of this phenomenon is the fact that the Hodge-Elliptic genus is not an index and in particular jumps at points in the moduli space where extra chiral currents appear. For example the generic Hodge-Elliptic genus should capture the physics of the infinite volume limit, while the orbifold Hodge-Elliptic genus is defined at a particular point where the orbifold $\torus^4/ \zed_2$ is resolved. Therefore one expects, and indeed finds, two distinct partition functions $\DT^{\mathrm{ref}}$. Less clear is the situation for the complex Hodge-Elliptic genus for which a certain SU(2) vertex operator algebra extends the $\cN=4$ superconformal algebra \cite{Creutzig:2013mqa}. Since this quantity is independent of complex structure deformations \cite{Kachru:2016igs,Wendland:2017eiw} and explicitly computable it is natural to expect that it captures the physics somewhere in the K3 moduli space. Unfortunately we couldn't find explicitly this locus or show its existence; at this stage this is only a conjecture as no deformation of the large radius sigma model is known which would lead to such a point. 

Geometrically we expect the BPS invariants to be defined in terms of the intersection theory of the reduced Hilbert scheme $\mathrm{Hilb}^{h,d,n} (X) / E$. Let us denote by $\mathscr{M}_\gamma$ the relevant moduli spaces for a charge vector $\gamma$. The general problem in extracting enumerative invariants from the spaces $\mathscr{M}_\gamma$ is that these are poorly understood with several singularities and branches of different dimensions. If $\mathscr{M}_\gamma$ were smooth one could define the refined Donaldson-Thomas invariants as $\chi_u$ genera of such moduli spaces, appropriately normalized. Following \cite{Kachru:2016igs}, it is natural to identify the invariants $\DT^{\mathrm{ref}}_{h,d,n} $ with motivic invariants. Roughly speaking one considers $\left[ \mathscr{M}_\gamma \right]$ as a class in the abelian K-theory group of varieties, generated by isomorphism classes of complex varieties modulo the scissor relations, extended by $\motive^{-1}$, the formal inverse of the Lefschetz motive (the class of the affine line). Taking the symmetrized Poincar\'e polynomials
\be
\DT^{\mathrm{ref}}_{h,d,n} = \mathrm{P} \left(\left[ \mathscr{M}_\gamma \right] \right) \in \mathbb{Q} [u , u^{-1}]
\ee
provides a working definition for the refined invariants.

It is tempting to speculate that such objects could more naturally be obtained in terms of the M2-brane index, defined in \cite{membranes} for smooth toric manifolds and in \cite{Cirafici:2021yda} in noncommutative chambers. Indeed it was shown in \cite{Bryan:2015uva} that the unrefined invariants with $h=0,1$ can be computed directly via toric localization. While the compact geometry is not toric, there exist in this case a stratification of the relevant moduli spaces where each strata is separately toric. Roughly speaking the strategy of \cite{membranes,Cirafici:2021yda} is to perform the localization computation by keeping all the toric weights and then identifying the product of all the toric weights with the refined parameter. Such product is kept finite in a scaling limit which sends all the weights to zero or to infinity. While in \cite{Bryan:2015uva} the toric action is different in each strata, one can speculate that a refined parameter introduced in this way should be identified across different strata, and would correspond to a square root of the canonical bundle of $\mathrm{K3} \times E$. We leave such speculations for future work.

Regrettably, speculations notwithstanding, we don't have a precise definition of the refined invariants nor we understand how to precisely associate a counting function with a point in the moduli space. Presumably this would entail introducing a theory of stability conditions on the reduced Hilbert scheme of $\mathrm{K3} \times E$. In this note we take a working approach and compute such invariants in the hope that this could help clarifying their geometrical meaning.

It's worth mentioning that certain refined BPS invariants can also be derived by duality constraints \cite{Alexandrov:2020qpb,Alexandrov:2019rth}. As already discussed in Section \ref{aspects} in order to compare with the geometrical results of \cite{Bryan:2015uva} we use an expansion in a different variable from what is commonly used in the literature. This give results which are not immediately comparable, although we hope to return to this problem in the future. Finally in \cite{Sen:2012hv} it was proposed to interpret the counting of refined invariants from a representation theory perspective as a refinement of a certain helicity supertrace.

Let us go back to the definition \eqref{PhirefDTref} and try to elucidate its structure. As in \cite{Kachru:2016igs} we have
\be
\sum_{k=0}^{\infty} p^k \, \HEll \left( \mathrm{Hilb}^{[k]} (\mathrm{K3}) \right) = \prod_{r>0,s\ge0,t,v} \frac{1}{(1 - q^r y^t p^r u^v)^{c(r\, s,t,v)}} = \frac{p \, \phi_{KKP} (\tau , z , \nu)}{\Phi^{\mathrm{ref}} (\tau , z , \nu, \sigma) } \, .
\ee
Note that 
\begin{align}
 \prod_{r>0,s\ge0,t,v} (1 - q^r y^t p^r u^v)^{c(r\, s,t,v)} =& \exp \sum_{r \ge 1} \sum_{s \ge 0} \sum_{t,v \in \zed} c(r\, s,t,v) \log \left( 1 - q^r y^t p^r u^v \right)  
 \cr
 =& \exp \sum_{r \ge 1} \sum_{s \ge 0} \sum_{t,v \in \zed} c(r\, s,t,v) \left( - \sum_{d=1}^\infty \frac{ q^{d r} y^{d t} p^{d r} u^{d v} }{d}  \right) 
 % \cr =& \exp \left( - \sum_{r \ge 1} \sum_{s \ge 0} \sum_{t,v \in \zed}  \sum_{d=1}^\infty c\left( d^2 \frac{r}{d} \, \frac{s}{d},\frac{d t}{d},\frac{d v}{d} \right)  \frac{ q^{d r} y^{d t} p^{d r} u^{d v} }{d}  \right)
\cr
 =& \exp \left( - \sum_{r \ge 1} \sum_{s \ge 0} \sum_{t,v \in \zed}  \sum_{d \vert (r,s,t,v)}^\infty \frac{1}{d} c \left( \frac{r s}{d^2},\frac{t}{d},\frac{v}{d} \right)   q^{r} y^{t} p^{r} u^{v}  \right) \, .
\end{align}

Equivalently 
\be
\Phi^{\mathrm{ref}} (\tau , z , \nu, \sigma) = p \, \phi_{KKP} (\tau , z , \nu) \, \exp \left( - \sum_{r=1}^\infty p^r  \left( \HEll \vert V_r \right) (\tau , z , \nu , \sigma ) \right)
\ee
where we have introduced the Hecke-like operators
\be
V_r \ : \ \sum_{s,t,v} c (s,t,v) q^r y^t u^v \longrightarrow  \sum_{s \ge 0} \sum_{t,v \in \zed}  \sum_{d \vert (r,s,t,v)}^\infty \frac{1}{d} c \left( \frac{r s}{d^2},\frac{t}{d},\frac{v}{d} \right)   q^{r} y^{t} u^{v} 
\ee
which implement the plethystic symmetrization. If we introduce the notation
\be
\zeta_r =  \left( \HEll \vert V_r \right) (\tau , z , \nu , \sigma ) 
\ee
we can then express 
\be
\HEll  (\mathrm{Hilb}^{[m]}) = s_m (\zeta_1 , \dots , \zeta_m)
\ee
in terms of the Schur polynomials (for the symmetric representations). The first few are
\begin{align}
s_1 (\zeta_1) &= \zeta_1 \, , \cr
s_2 (\zeta_1 , \zeta_2) & = \frac12 \zeta_1^2 + \zeta_2 \, , \cr
s_3 (\zeta_1 , \zeta_2 , \zeta_3) & = \frac16 \zeta_1^3 + \zeta_1 \zeta_2 + \zeta_3 \, .
\end{align}
Explicitly now
\begin{align}
\HEll \left( \mathrm{Hilb}^{[1]} (\mathrm{K3}) \right) & = \HEll (\tau , z , \nu) \, , \\
\HEll \left(\mathrm{Hilb}^{[2]} (\mathrm{K3}) \right) &= \frac12 \HEll (\tau , z , \nu)^2 + \frac12 \left[ \HEll (\frac{\tau}{2} , z , \nu) + \HEll (2 \tau , 2 z , 2 \nu) + \HEll (\frac{\tau+1}{2} , z , \nu)\right] \, , \cr
\HEll \left(\mathrm{Hilb}^{[3]} (\mathrm{K3}) \right) & = \frac16 \HEll (\tau , z , \nu)^3 
\cr & + \frac12 \HEll (\tau , z , \nu)  \left[ \HEll (\frac{\tau}{2} , z , \nu) + \HEll (2 \tau , 2 z , 2 \nu) + \HEll (\frac{\tau+1}{2} , z , \nu) \right] \cr
& + \frac13 \left[
\HEll (\frac{\tau}{3} , z , \nu) + \HEll (3 \tau ,3 z , 3 \nu)  + \HEll (\frac{\tau+1}{3} , z , \nu)  + \HEll (\frac{\tau+2}{3} , z , \nu) 
\right] \nonumber \, .
\end{align}
In particular, comparing term to term
\begin{align} \label{ExpansionDTHEll}
 \DT^{\mathrm{ref}} & =  \frac{1}{p} \DT^{\mathrm{ref}}_0 + \DT^{\mathrm{ref}}_1 + p \, \DT^{\mathrm{ref}}_2 + \cdots  
\\ \nonumber
& = - \frac{1}{p} \frac{1}{\phi_{KKP} (\tau , z, \nu)}
\left(  \HEll \left( \mathrm{Hilb}^{[0]} (\mathrm{K3}) \right)  + p \,  \HEll \left( \mathrm{Hilb}^{[1]} (\mathrm{K3}) \right)  + p^2 \,  \HEll \left( \mathrm{Hilb}^{[2]} (\mathrm{K3}) \right) + \cdots \right)
\end{align}
we see that we can write explicitly the refined invariants in terms of the Hodge-Elliptic genus and its powers (computed at different values of the arguments)
\begin{align}
\DT^{\mathrm{ref}}_0 &= - \frac{1}{\phi_{KKP} (\tau , z , \nu)} \, , \cr
\DT^{\mathrm{ref}}_1 &=  \DT^{\mathrm{ref}}_0 \,  \HEll (\tau , z , \nu) \, , \cr
\DT^{\mathrm{ref}}_2 &=  \DT^{\mathrm{ref}}_0 \,  \frac12 \left[ \HEll (\tau , z , \nu)^2 +  \HEll (\frac{\tau}{2} , z , \nu) + \HEll (2 \tau , 2 z , 2 \nu) + \HEll (\frac{\tau+1}{2} , z , \nu)\right]  \, .
\end{align}
Using the explicit form of the Hodge-Elliptic genus we have a prediction for the full series of refined Donaldson-Thomas invariants, order by order. Now we will specialize to different version of the Hodge-Elliptic genus and try to extract geometrical information from the counting functions.

The first prediction was actually proven in \cite{Katz:2014uaa}
\be \label{DT0ref}
\DT^{\mathrm{ref}}_0 = - \frac{1}{\phi_{KKP} (\tau , z , \nu)} = \frac{1}{\eta(\tau)^{18} \ \theta_1 (\tau , z + \nu) \ \theta_1 (\tau , z-\nu)}
\ee
and is universal, independent of the particular Hodge-Elliptic genus one is considering.

Consider first the complex Hodge-Elliptic partition function \eqref{HEll1}. Multiplying it by \eqref{DT0ref}  and using the identities \eqref{diag12} and \eqref{vertmu} we find the prediction
\begin{align} \label{DT1refc}
\DT_1^{ref,c} &= - \left( \frac{\theta_1 (q,y)^2}{\theta_1 (q , u y) \, \theta_1 (q , \frac{y}{u})} \right) \frac{1}{\eta(q)^{24}} \Big[ \left( 20 + 2  u + \frac2u  \right) \left( 1+
 \frac{y}{(1-y)^2} + \sum_{d=1}^\infty \, \sum_{k \vert d} k (y^k + y^{-k}) \, q^d \right) \nonumber \\ 
 &  - \left( 18 + 2 u+ \frac{2}{u} \right)  + 24 \sum_{d=1}^\infty \, \sum_{k \vert d} k (- 2 ) \, q^d  + \left( 2 - \frac1u - u \right) \left( \frac{q^4 + 2 q^3 + 6 q^2 + 2 q + 1}{(q-1) (q+1) (q^2+ q + 1)}  \right)
\Big] \, .
\end{align}
One of the effects of the refinement is to produce an overall ratio of theta functions. We see that the vertical curves with nodal fibers are refined also by a multiplicative factor $\left( 20 + 2 u + \frac2u  \right)$, which replaces the 24 which counts the number of singular fibers. This factor is replaced by $\left( 18 + 2 u+ \frac{2}{u} \right) $ for smooth vertical curves. The refinement of the diagonal curves is more complicated. The contribution from the diagonal curves captured by the Eisenstein series should really be paired with the rational function, since they are the only $q$-dependent terms without any $y$-dependence (up to the constant term in the expansion of the rational function). Indeed we show in Appendix \ref{cyclo} that the rational function can be interpreted as shifting the first terms in the Eisenstein sum, when this is rephrased in terms of cyclotomic polynomials. However we don't have a simple geometrical interpretation of this.

Consider now the Kummer surface, the $\zed_2$ orbifold of complex two-tori. In this case we consider the orbifold Hodge-Elliptic partition function \eqref{HEllorb} multiplied by \eqref{DT0ref}. By using again the identities \eqref{diag12} and \eqref{vertmu} we find the prediction
%\begin{align} \label{DT1reforb}
%\DT_1^{\mathrm{ref,orb}} &=  - \left( \frac{\theta_1 (q,y)^2}{\theta_1 (q , u y) \, \theta_1 (q , \frac{y}{u})} \right) \frac{1}{\eta(q)^{24}} \Big[ \left( 20 + 2 \left( u + \frac1u \right) \right) \left( 
% \frac{y}{(1-y)^2} + \sum_{d=1}^\infty \, \sum_{k \vert d} k (y^k + y^{-k}) \, q^d \right) \nonumber \\ 
%& + \left( \frac{1}{u} + u \right) + (28 -\frac{2}{u}- 2 u) \left(  \sum_{d=1}^\infty \, \sum_{k \vert d} k (- 2 ) \, q^d \right) - \left( 16 - \frac{8}{u} - 8 u \right) \left( \sum_{d=1}^\infty \, \sum_{k \vert d} k (- 2 ) \, q^{2 d} \right) \Big]
%\end{align} 
\begin{align} \label{DT1reforb}
\DT_1^{\mathrm{ref,orb}} &=  - \left( \frac{\theta_1 (q,y)^2}{\theta_1 (q , u y) \, \theta_1 (q , \frac{y}{u})} \right) \frac{1}{\eta(q)^{24}} \Big[ \left( 20 + 2  u + \frac2u  \right) \left( 1 + 
 \frac{y}{(1-y)^2} + \sum_{d=1}^\infty \, \sum_{k \vert d} k (y^k + y^{-k}) \, q^d \right) \nonumber \\ 
& \!\!\!\!\!\!\!\! \!\!\!\!\!\!\!\!  - \left( 20 + u + \frac1u \right) + \left( 28 -\frac{2}{u}- 2 u \right) \left(  \sum_{d=1}^\infty \, \sum_{k \vert d} k (- 2 ) \, q^d \right) - \left( 16 - \frac{8}{u} - 8 u \right) \left( \sum_{d=1}^\infty \, \sum_{k \vert d} k (- 2 ) \, q^{2 d} \right) \Big]
\end{align} 
We see that the contribution of the vertical curves with nodal fibers has the same structure as in the previous case \eqref{DT1refc}, while now smooth vertical curves contribute with a weight $\left( 20 + u + \frac1u \right)$. More complicated is the situation for diagonal curves. Similarly to what happens in \eqref{DT1refc} the effect of the refinement on diagonal curves involves correcting the coefficients of the Eisenstein series by $u$-dependent shifts. As above it is not clear what is the geometrical interpretation.

%In this case we write the Hodge-Elliptic genus as
%\begin{align}
%\HEll^{\mathrm{orb}} (\tau , z , \nu) = & \frac{1}{24} \left(\frac{2}{u} + 20 + 2 u \right) \frac{\theta_1 (\tau , z)^2}{\eta(\tau)^3}  \left[ 24 \frac{2 F_2^{(2)}(\tau)}{\eta (\tau)^3} + 24 \mu (\tau , z) \right] + 
%\left( 1 - \frac{1}{2 u} - \frac{u}{2} \right) \frac{\theta_1 (\tau , z)^2}{\eta (\tau)^6} 
%\cr
%& +  \frac{\theta_1 (\tau , z)^2}{\eta(\tau)^3} \frac{E_2 (\tau)}{\eta (\tau)^3} \left( - \frac73 + \frac{1}{6 u} + \frac{u}{6} \right) + \frac43 \left(1 - \frac{1}{2 u} - \frac{u}{2} \right) \frac{\theta_1 (\tau , z)^2}{\eta(\tau)^3} \frac{E_2 (2 \tau)}{\eta (\tau)^3}
%\end{align}
%Using the identity \eqref{} and \eqref{} we write
%\begin{align}
%\HEll^{\mathrm{orb}} (\tau , z , \nu) = & \frac{1}{24} \left(\frac{2}{u} + 20 + 2 u \right) \frac{\theta_1 (\tau , z)^2}{\eta(\tau)^6} 24 \left( 
% \frac{y}{(1-y)^2} + \sum_{d=1}^\infty \, \sum_{k \vert d} k (y^k + y^{-k}) \, q^d \right) \cr & + 
%\left( 1 - \frac{1}{2 u} - \frac{u}{2} \right) \frac{\theta_1 (\tau , z)^2}{\eta (\tau)^6} 
%+   \frac{1}{12}  \frac{\theta_1 (\tau , z)^2}{\eta(\tau)^6} \left( - \frac73 + \frac{1}{6 u} + \frac{u}{6} \right)\left( 12 +  \sum_{d=1}^\infty \, \sum_{k \vert d} k (- 2 ) \, q^d \right)
%\cr &
%+ \frac{1}{9}  \left(1 - \frac{1}{2 u} - \frac{u}{2} \right) \frac{\theta_1 (\tau , z)^2}{\eta(\tau)^6} \left( 12 +  \sum_{d=1}^\infty \, \sum_{k \vert d} k (- 2 ) \, q^{2 d} \right)
%\end{align}
%Now multiplying by \eqref{DT0ref} we find
%\begin{align}
%\DT_1^{\mathrm{ref,orb}} =  
%\end{align} 

Finally let us consider the generic Hodge-Elliptic genus \eqref{genericHEll2}. 
%We can write it as
%\begin{align}
%\HEll^{g} (\tau , z, \nu) 
%%&= \mathsf{Ell} (\tau , z) + \left( 2 - \frac{1}{u} - u \right) \frac{\theta_1 (\tau , z)^2}{\eta (\tau)^3} \left[ q^{-\frac18} - 2 \mu (\tau , z) \right] 
%%\cr &
%=  \frac{1}{24} \left( 20 + \frac{2}{u} + 2 u \right) \mathsf{Ell} (\tau , z) + \left( 2 - \frac{1}{u} - u \right) \frac{\theta_1 (\tau , z)^2}{\eta (\tau)^3} \left[ \frac{1}{12} H (\tau) + q^{-\frac18} \right]
%\end{align}
As before multiplying by \eqref{DT0ref} we obtain
%\begin{align} \label{DT1refg}
%\DT_1^{\mathrm{ref,g}} &=  - \left( \frac{\theta_1 (q,y)^2}{\theta_1 (q , u y) \, \theta_1 (q , \frac{y}{u})} \right) \frac{1}{\eta(q)^{24}} 
% \left( 20 + 2  u + \frac{2}{u}  \right) \left( \frac{1}{12} + 
% \frac{y}{(1-y)^2} + \sum_{d=1}^\infty \, \sum_{k \vert d} k (y^k -2 +y^{-k}) \, q^d \right) \nonumber \\ 
%& +  \left( \frac{\theta_1 (q,y)^2}{\theta_1 (q , u y) \, \theta_1 (q , \frac{y}{u})} \right) \frac{1}{\eta(q)^{22}}  \left( 2 - \frac{1}{u} - u \right) \frac{\theta_1 (\tau , z)^2}{\eta (\tau)^3} \left[ \frac{1}{12} H (\tau) + q^{-\frac18} \right]
%\end{align} 
\begin{align} \label{DT1refg}
\DT_1^{\mathrm{ref,g}} &=  - \left( \frac{\theta_1 (\tau ,z)^2}{\theta_1 (\tau , z+\nu) \, \theta_1 (\tau , z-\nu)} \right) \frac{1}{\eta(\tau)^{24}} 
\Bigg[ - \frac{22}{24} \left( 20 + \frac2u + 2 u \right) 
   \nonumber \\ 
&
+\left( 20 + \frac2u + 2 u \right) \sum_{d=1}^\infty \, \sum_{k \vert d} (-2 k) \, q^d 
+ \left( 20 + 2  u + \frac{2}{u}  \right) \left( 1 + 
 \frac{y}{(1-y)^2} + \sum_{d=1}^\infty \, \sum_{k \vert d} k (y^k +y^{-k}) \, q^d \right) 
 \nonumber \\ 
%& + \frac{22}{24} \left( 20 + \frac2u + 2 u \right) - 2  \left( 20 + \frac2u + 2 u \right) \sum_{d=1}^\infty \, \sum_{k \vert d} k \, q^d 
%   \nonumber \\ 
&
 -  \left( 2 - \frac{1}{u} - u \right) \eta^3 (\tau) \left( \frac{1}{12} H (\tau) + q^{-\frac18} \right) \Bigg]
\end{align} 
The first two lines is a simple multiplicative refinement of the Elliptic genus and we recognize immediately the contributions of the various types of curves. The last line is more challenging to interpret. Also in the two previous cases one can start with a multiplicative refinement of the Elliptic genus, as in \eqref{HEll2} or \eqref{HEllorb}, but then the extra terms conspire precisely to alter the weights of the contributions of various curves. Here the situation is different, due to the appearance of the function $H (\tau)$. As we have remarked in Section \ref{DTK3xE} this function has a clear interpretation in terms of the geometry of $\mathrm{K3} \times E$ only when in combination with the Appell-Lerch sum $\mu (\tau , z)$. Its appearance in the refined Donaldson-Thomas generating function leads to the surprising prediction that in the refined setting its coefficients, which are related to the dimensions of the irreducible representations of the Mathieu group $\mathbb{M}_{24}$, should have a direct geometric representation in terms of the reduced Hilbert scheme on $\mathrm{K3} \times E$. Note that this does not happen in $\phi_{KKP} (\tau , z , \nu)$ but only in the higher order terms in the Donaldson-Thomas expansion. Furthermore the term $(2-\frac{1}{u} - u)$ guarantees that the unrefined theory is blind to the action of $\mathbb{M}_{24}$ on the Hilbert space of BPS states. 

Consider now the Hilbert space decomposition $\cH_{h=1} = \bigoplus_{d,n} \cH_{d,n} q^{d-1} (-y)^{n}$ where the Donaldson-Thomas invariant is interpreted as a refined Witten index associated to each Hilbert space subfactor. Each $\cH_{d,n}$ can be in turn decomposed as a sum of factors according to the structure of \eqref{DT1refg} and \eqref{DT0ref}, for example by keeping track of the diagonal and vertical curve contributions. Due to the presence of the function $H(\tau)$ in \eqref{DT1refg}, we see that one of these factors carries the action of the Mathieu group. It would be interesting to develop this further.

Higher order enumerative invariants can be computed similarly, using the expansion \eqref{ExpansionDTHEll}. The first few orders for $\DT_2^{\mathrm{ref}}$ and $\DT_3^{\mathrm{ref}}$ for all three Hodge-Elliptic genera can be found in the supporting \textsc{mathematica} file \cite{mathfile}.

\section{The Mathieu group and twined complex Hodge-Elliptic genera} \label{Mathieutwined}

%look at hohn creutzig for the Matheiu groups BE CAREFUL ABOUT THE SIGNS, is there an overall minus? is M23 also called mathieu?

%also look at chapter 6.3 twined elliptic genera by localization of the thesis by Kidambi automorphic forms in string theory.

%The computation will probably work better when the integral of the elliptic genus is expressed in terms of the characters of the bundle, so that only the trivial character has to be localized

% from cheng anagiannis review

In this Section we will consider the case where there is a finite symmetry group $G$ which commutes with the superconformal symmetries of the model. In this case one can defined the elliptic genus twined by $g \in G$, see \cite{Gaberdiel:2010ch,Gaberdiel:2011fg,Cheng:2016org,Eguchi:2010fg} for a sample of the literature. 
%We will compute several examples of the twined complex Hodge-Elliptic genera.
%\be
%\mathsf{Ell}_g = \Tr_{\cH_{RR}} \left(
%g (-1)^{F_L + F_R} y^{F_L} q^{L_0 - \frac{c}{24}} \overline{q}^{\overline{L}_0 - \frac{c}{24}}
%\right)
%\ee
%\subsection{Twined elliptic genera and symplectic automorphisms of K3 surfaces}
We can generalize \eqref{EllCFT} and define the elliptic genus twined by any $g \in G$ as
\be
\mathsf{Ell}_g = \Tr_{\cH_{RR}} \left(
g \,  (-1)^{J_0 + \overline{J}_0} y^{J_0} q^{L_0 - \frac{c}{24}} \overline{q}^{\overline{L}_0 - \frac{c}{24}} 
\right) \, .
\ee
In this Section we will introduce twined complex Hodge-Elliptic genera and compute explicitly several examples.

%\subsection{Symplectic automorphisms of K3 surfaces}

In particular there is a class of twining genera which are induced by certain geometric automorphisms of K3 surfaces. A geometric automorphism which acts trivially on the holomorphic 2-forms of the K3 is called symplectic. The automorphisms which have finite order, so that they only have isolated fixed points, are classified and correspond to certain subgroups of the Mathieu group $\mathbb{M}_{23}$, with certain conditions. The non-trivial minimal subgroups are given in the following table, together with their \textsc{atlas} names as elements of $\mathbb{M}_{24}$, the number of fixed points and their eigenvalues at the fixed point:
%\begin{equation} \label{tableFP}
%\begin{array}{|c|c|c|}
%\hline
%\text{conjugacy class} & \text{number of fixed points} & \text{eigenvalues at the fixed points} \cr \hline
%\mathrm{2A} & 8 & 8 \times (-1,-1) \cr
%\mathrm{3A} & 6 & 6 \times (\zeta_3 , \zeta_3^{-1}) \cr
%\mathrm{4B} & 4 & 4 \times (\ii , - \ii) \cr
%\mathrm{5A} & 4 & 2  \times (\zeta_5 , \zeta_5^{-1}) \ \text{and}  \ 2  \times (\zeta_5^2 , \zeta_5^{-2})  \cr
%\mathrm{6A} & 2 & 2 \times (\zeta_6 , \zeta_6^{-1}) \cr
%\mathrm{7AB} & 3 & (\zeta_7 , \zeta_7^{-1}), (\zeta_7^2 , \zeta_7^{-2}) \ \text{and} \ (\zeta_7^3 , \zeta_7^{-3}) \cr
%\mathrm{8A} & 2 &  (\zeta_8 , \zeta_8^{-1}) \ \text{and}  \ (\zeta_8^3 , \zeta_8^{-3}) \cr
%\hline
%\end{array}
%\end{equation}

\begin{table}[h] \label{FPtable}
\begin{tabular}{|c|c|c|c|}
\hline
conjugacy class & number of fixed points & eigenvalues at the fixed points & $\tilde{T}_g (\tau)$ \cr \hline
$\mathrm{2A}$ & 8 & $8 \times (-1,-1) $ & $ 16 \Lambda_2$ \cr
$\mathrm{3A}$ & 6 & $6 \times (\zeta_3 , \zeta_3^{-1})$  & $ 6 \Lambda_3$ \cr
$\mathrm{4B}$ & 4 & $4 \times (\ii , - \ii) $ & $ 4 \left( - \Lambda_2 + \Lambda_4 \right)$ \cr
$\mathrm{5A}$ & 4 & 2  $\times (\zeta_5 , \zeta_5^{-1}) \ \text{and}  \ 2  \times (\zeta_5^2 , \zeta_5^{-2})$  & $ 2 \Lambda_5$ \cr
$\mathrm{6A}$ & 2 & 2 $ \times (\zeta_6 , \zeta_6^{-1})$ & $2 \left( - \Lambda_2 - \Lambda_3 + \Lambda_6 \right)$ \cr
$\mathrm{7AB}$ & 3 & $(\zeta_7 , \zeta_7^{-1}), (\zeta_7^2 , \zeta_7^{-2}) \ \text{and} \ (\zeta_7^3 , \zeta_7^{-3}) $ & $ \Lambda_7$ \cr
$\mathrm{8A}$ & 2 &  $(\zeta_8 , \zeta_8^{-1}) \ \text{and}  \ (\zeta_8^3 , \zeta_8^{-3})$ & $ -\Lambda_4 + \Lambda_8$ \cr
\hline
\end{tabular}
\caption{Data characterizing the non-trivial minimal subgroups and their $\tilde{T}_g (\tau)$ functions.} \label{tableFP}
\end{table}

The corresponding Elliptic genus can be computed as before, with the result
\begin{equation}
\mathsf{Ell}_g (\tau , z)  = \frac{1}{12} \chi_g (\mathrm{K3}) \, \varphi_{0,1} (\tau , z) + \tilde{T}_g (\tau) \, \varphi_{-2,1} (\tau , z)
\end{equation}
where $\chi_g (\mathrm{K3})$, counting the number of fixed points, and the function $ \tilde{T}_g (\tau)$ can be read off table \ref{tableFP}. The functions $\Lambda_N (\tau)$ were introduced in \eqref{Lambdadef}.
%\be
%\Lambda_N (\tau) = N q \frac{\dd}{\dd q} \left( \log \frac{\eta (N \tau)}{\eta (\tau)}\right)
%\ee
%is a weight two modular form on $\Gamma_0 (N)$.

This result follows from the fixed point formula applied to the integral \eqref{EllFrep} 
%\be
%\mathsf{Ell} (\tau , z) = \sum_{N=0}^\infty \mathscr{F}_N \int \ch (S^N \cT) \mathrm{Todd} (\cT)
%\ee
by writing
\be
% \int \chi_N (\exp{x}) \mathrm{Todd} (\e^x , \e^{-x}) 
 \int \ch (S^N \cT) \mathrm{Todd} (\cT)
= \sum_{\lambda_i} \frac{ \chi_N (\lambda_i) }{\left( 1 - \lambda_i \right)\left( 1- \lambda_i^{-1} \right)} \, ,
\ee
where the eigenvalues $\lambda_i$ can be read off table \eqref{tableFP}, and $\chi_N$ denotes the SU(2) characters. 
%Then the computation goes as follows... (sketch general computation for twining genera? doing it in detail is to complicated since different $h$ structure for each case)
%Recall that the function $\mathscr{F}$ is defined as
%\be
%\mathscr{F}_N =- \frac{\theta_1 (\tau,z)}{\eta(\tau)^3} \left[ s_N (\tau) - 2 s_{N+1} (\tau) + 2 s_{N+3} (\tau) - s_{N+4} (\tau) \right]
%\ee
%with
%\be
%s_N (\tau) = \frac{\theta_1 (\tau ,z)}{\eta(\tau)^3} \widetilde{s}_N (\tau) + \delta_{N,1} \theta_1 (\tau,z) \, \mu (\tau,z)
%\ee
%with
%\be
%\widetilde{s}_N (\tau) = \left\{
%\begin{matrix}
%\frac{q}{1-q} & \text{if} \ N=0 \\
%F_2^{(2)} (q) & \text{if} \ N=1 \\
%\frac{N-1}{1-q^{N-1}} & \text{otherwise} 
%\end{matrix}
%\right.
%\ee
The result can be compactly written as an equivariant Euler characteristic
\be \label{equivEuler}
\mathsf{Ell}_g (\tau , z) = \sum_{N=0}^\infty \mathscr{F}_N  (\tau, z) \sum_{\lambda_i} \frac{ \chi_N (\lambda_i) }{\left( 1 - \lambda_i \right)\left( 1- \lambda_i^{-1} \right)} \, ,
\ee
where the function $\mathscr{F}_N (\tau, z)$ was defined in \eqref{functionF} in terms of the functions \eqref{functions} and \eqref{functionstilde}. The computation proceeds as in Section \ref{Ellrevisited}. For example in the case $[g] = 2A$
\begin{align}
\mathsf{Ell}_{[g] = 2A} (\tau , z) = \frac{\theta_1 (\tau , z)^2}{\eta (\tau)^3} \left[ \frac{1}{\eta (\tau)^3} \left( - 2 \widetilde{s}_0 (\tau) + 8 \widetilde{s}_1 (\tau) - 14 \widetilde{s}_2 (\tau) + 16 \sum_{N=3}^\infty (-1)^{N+1}  \widetilde{s}_N (\tau) \right) + 8 \mu (\tau ,z) \right] \, .
\end{align}
The quantity in round brackets can be written as
\begin{align}
8 \widetilde{s}_1 (\tau) + \frac13 \left( -2 \widetilde{s}_0 (\tau) + 50 \widetilde{s}_2 (\tau) + 48 \sum_{N=3}^\infty \widetilde{s}_N (\tau) \right) - \frac13 \left( 
4 \widetilde{s}_0 (\tau) + 92 \widetilde{s}_2 (\tau) + 96 \sum_{N=2}^\infty \widetilde{s}_{2 N} (\tau) \right)
\cr = \frac13 (-2 E_2 (\tau) + 48 F_2^{(2)} (\tau)) - 16 \Lambda_2 (\tau)
\end{align}
where we have repeatedly used the analytical continuation\footnote{We stress again that the above identities do not hold in a truncated form.} $\sum_{n=1}^\infty n = - 1/12$ and \eqref{E2ancont} to write
\begin{align}
16 \Lambda_2 (\tau) & = \frac43 \left( -1 -48 \sum_{n=1}^\infty \frac{n}{1-q^{2n}} + 24 \sum_{n=1}^\infty \frac{n}{1-q^n} \right) = \frac43 \left( -1 + \frac{24}{1-q} + 24 \sum_{N=1}^\infty \frac{2N -1}{1-q^{2N-1}} \right) \nonumber
\\ & = \frac13 \left( 4 \widetilde{s}_0 (\tau) + 92 \widetilde{s}_2 (\tau) + 96 \sum_{N=2}^\infty \widetilde{s}_{2 N} (\tau) \right) \, .
\end{align}
By putting everything together we can write
\begin{align}
\mathsf{Ell}_{[g] = 2A} (\tau , z) 
 = \frac13 \mathsf{Ell} (\tau , z) - 16 \Lambda_2 (\tau) \frac{\theta_1 (\tau  ,z)^2}{\eta (\tau)^6}
%\cr
= \frac{\chi_{[g]=2A} (\mathrm{K3})}{12} \varphi_{0,1} (\tau , z) + 16 \Lambda_2 (\tau) \varphi_{-2,1} (\tau , z) \, .
\end{align}
The remaining cases can all be treated similarly and are left to the reader.

%\textbf{but write down how the lambda appears from the s?}

%\subsection{Twining Hodge-Elliptic genera}

We can now define a twining Hodge-Elliptic genus as
\be
\HEll_g (\tau , z , \nu) = \Tr \left( g \, (-1)^{J_0 + \overline{J}_0} y^{J_0} u^{\overline{J}_0} q^{L_0 - \frac{c}{24}} \right)
\ee
where as before the right movers are in the ground states. Geometrically we can start from the relation \eqref{HEll-Ell-Euler} to define
\begin{align}
\HEll_g (\tau , z , \nu) = \mathsf{Ell}_g (\tau , z)  - \left[ 1 - \frac12 \left( u + \frac{1}{u} \right) \right] \mathscr{F}_0 (\tau, z)  \, \chi_g (\cO) \, .
\end{align}
The computation of the complex Hodge-Elliptic genus proceeds as in Section \ref{complexHEll}, by weighting the contribution of the flat bundles with an appropriate factor. The flat bundle contribution is the $N=0$ term in the sum \eqref{equivEuler}. One can see by direct computation that this term is the same for all conjugacy classes and is determined by the combination
\be
-2 \widetilde{s}_0 (\tau) + 4  \widetilde{s}_1 (\tau) -4  \widetilde{s}_3 (\tau) + 2  \widetilde{s}_4 (\tau) \, .
\ee 
Then repeating the computations of Section \ref{complexHEll} step by step now gives the flat bundle contribution
\begin{align}
\mathsf{Flat} (\tau,z) = \frac12 \frac{\theta_1 (q,y)^2}{\eta(q)^3}  8 \mu (q,y) + \frac{\theta_1 (q,y)^2}{\eta(q)^6} \left[
\frac{2 (1 + 2 \, q + 6 \, q^2 + 2 \, q^3 + q^4)}{(q-1)(q+1)(1+q+q^2)} + 8 F_2^{(2)} (q)
\right]
\end{align}
so that the Hodge-Elliptic twining genera are given by
\begin{equation}
\mathsf{HEll}_g (\mathrm{K3})  = \frac{1}{12} \chi_g (\mathrm{K3}) \, \varphi_{0,1} + \tilde{T}_g \, \varphi_{-2,1} - \left( 1 - \frac12 \left( u + \frac{1}{u} \right) \right) \, \mathsf{Fl} (q,y)
\end{equation}
Equivalently we can rewrite this as
\begin{align}
\mathsf{HEll}_g (\mathrm{K3})  = & \frac{1}{24} \chi_{g ,u} (\mathrm{K3}) \mathsf{Ell}_g (\tau , z) +  \tilde{T}_g \, \varphi_{-2,1} 
\cr &
- 2 \left( 1 - \frac12 \left(u + \frac{1}{u} \right) \right) \frac{\theta_1 (\tau , z)^2}{\eta (\tau)^6} \left[ \frac{(1 + 2 \, q + 6 \, q^2 + 2 \, q^3 + q^4)}{(q-1)(q+1)(1+q+q^2)} + \frac16  E_2 (\tau)
\right]
\end{align}
in terms of the equivariant $\chi_u$ genus
\be
 \chi_{g , u} (\mathrm{K3}) = \frac{1}{u}   \sum_{\lambda_i} \frac{ \left( 1 - u \lambda_i \right)\left( 1- u \lambda_i^{-1} \right) }{\left( 1 - \lambda_i \right)\left( 1- \lambda_i^{-1} \right)} =  2 u + \chi_g (\mathrm{K3}) - 4 + \frac{2}{u}
\ee
evaluated by localization.

\section{Conclusions} \label{Conclusions}

In this note we have studied the implications of three Hodge-Elliptic genera for the refined enumerative geometry of $\mathrm{K3} \times E$. We have found explicit formulas showing how the ordinary Donaldson-Thomas sums are modified, in the process elucidating also certain aspects of the ordinary Elliptic genus. The results should function as a starting point to study refinements of Donaldson-Thomas theory on $\mathrm{K3} \times E$. We conclude with a few comments and open problems:
\begin{itemize}
\item It would be very interesting to reproduce the results obtained here via a direct localization computation. As already mentioned in the text, a possible strategy is to use the full K-theory vertex and afterwards take a certain scaling limit of the toric weights. A reason this could work is that the localization computation of the ordinary enumerative invariants is carried on by using a stratification of the moduli space, where strata are separately toric. While the toric action will not glue globally, the scaling limit should individuate a parameter associated with the square root of the canonical bundle, and therefore should make sense globally. Note that there are several possible scaling limits of the toric weights, a fact perhaps explaining why different versions of the Hodge-Elliptic genus can appear.
\item We don't really understand the modular properties of Hodge-Elliptic genera. They fail to be modular but in a very specific way. It would be interesting to explore this issue further.
\item The relation between jumping phenomena, wall-crossing and the analytic properties of the refined BPS counting function should be established precisely.
\item An intriguing aspects of the construction is the appearance of the function $H(\tau)$ whose coefficients are related to the representations of the Mathieu group $\mathbb{M}_{24}$. This function appears in the generic Hodge-Elliptic genus and therefore in the enumerative geometry of $\mathrm{K3} \times E$. In contrast with what happens in the case of the Elliptic genus, in the refinement this function appears unpaired with the Appell-Lerch sum $\mu (\tau , z)$. In the unrefined case the combination of these two functions decomposes in a non trivial way to capture the geometry of curves in the target space. Unfortunately in this decomposition the individual coefficients are mixed up and all information of the representations of $\mathbb{M}_{24}$ is lost. Apparently this is not the case for the generic Hodge-Elliptic genus. Our results imply that it should be possible to see the Mathieu Moonshine phenomenon geometrically when one considers refined invariants which have non zero indices in both the K3 and the elliptic curve directions. This phenomenon appears only starting from $\DT_1$ but persists at higher orders. 
\item As we have explained, it is natural to conjecture that the refined partition functions that we have constructed in the course of the paper are related to a refinement of Donaldson-Thomas theory. This leads naturally to the further conjecture that the refined counting of BPS states is related to the refined topological string. As was conjectured in \cite{OP} and then proven in \cite{Oberdieck:2017zit}, the BPS counting function $\Phi_{10}$ can be expresses in terms of the (disconnected, reduced) Gromov-Witten invariants. 

%The Gromov-Witten theory of $K3 \times E$ can be defined as follows. Ordinary Gromov-Witten invariants of $\mathrm{K3} \times E$ vanish by deformation invariance; however one can define a reduced perfect obstruction theory by removing part of the obstruction bundle. The resulting reduced virtual fundamental class leads to the definition of reduced Gromov-Witten invariants
%\be
%\mathsf{GW}_{g,h,d}^\bullet = \int_{\left[ \overline{M}_{g,1}^\bullet (\beta, d) \right]^{red}} \mathrm{ev}^*_1 \left(  p_1^* (\beta^\vee) \cup p_2^* (\mathfrak{p})
%\right)
%\ee
%where $\mathfrak{p} \in H^2 ( E ; \zed)$ is the class Poincar\'e dual to a point and $\beta$ a primitive class such that $\langle \beta , \beta \rangle = 2 h -2$. Here $\overline{M}_{g}^\bullet (\beta, d) $ is the compactified moduli space of (possibly disconnected) genus $g$ curves to $\mathrm{K3} \times E$ representing the class $(\beta, d)$. Such a moduli space is acted upon by the elliptic curve $E$ with one dimensional orbits, hence the need to fix a point. 

One can define the generating function
\be
\mathsf{GW} = \sum_{g \in \zed} \sum_{h \ge 0} \sum_{d \ge 0} \mathsf{GW}_{g,h,d} \ \lambda^{2 \, g-2} p^{h-1} q^{d-1}
\ee
where the labels have the same meaning as in \eqref{DTgenfun}. The generating function of Gromov-Witten invariants is related to the Igusa cusp form as
\be
\mathsf{GW} = - \frac{1}{\Phi_{10}}
\ee
where the genus counting parameter $\lambda$ appears in the right hand side via the substitution $q = - \e^{\ii \lambda}$.

It is natural to propose that the GW-DT correspondence also holds in the refined setting for $\mathrm{K3} \times E$. This corresponds to considering the $\cN=4$ string on the $\Omega$-background and taking an appropriate limit \cite{membranes}. That such a construction should be possible is suggested by the fact that in the Donaldson-Thomas side the relevant moduli space admits a stratification where each strata admits a toric action, even if this does not glue globally. Since the GW-DT correspondence should hold in the large volume limit, it should be compared with the chiral Hodge-Elliptic genus.
As an aside comment, it is an interesting open problem to define the worldsheet theory of the refined $\cN=4$ topological string, generalizing the construction of \cite{Antoniadis:2006mr}.
\item We have computed certain twining complex Hodge-Elliptic genera associated to geometrical automorphisms of K3 surfaces. It would be interesting to extend this computation to all the available twisted-twined genera, as well as the generic and orbifold Hodge-Elliptic genera, to predict enumerative invariants which would refine the partition functions computed in \cite{Bryan:2018nlv,Chattopadhyaya:2017ews,David:2006ud,Gaberdiel:2012gf}

\end{itemize}

\section*{Acknowledgements}
I thank S. Alexandrov, P.A. Grassi, M. Kool, S. Mozgovoy and B. Pioline for discussions. These results were presented at the \textit{Workshop on Superstrings and Supermoduli}, SISSA 27-30 June 2022, and \textit{BPS states, Mirror Symmetry and Exact WKB II}, University of Sheffield, 5-9 September 2022. I thank the participants for discussions and the organizers for providing a cozy atmosphere and financial support. I am a member of INDAM-GNFM and I am supported by INFN via the Iniziativa Specifica GAST.

\section*{Data Availability Statement}

All data generated or analysed during this study are included in this article (and its
supplementary information files). Alternatively they are freely available for download at \url{https://cirafici.dmg.units.it/HEsupporting.nb}, as well as with the arxiv submission.

\section*{Conflict of Interest Statement}

The corresponding author states that there is no conflict of interest.

\appendix

\section{Some technical identities} \label{sidentities}

In this appendix we prove some of the technical identities used in the main text

\begin{lemma}
The following identities hold
\begin{align}
\sum_{i=-k}^k \frac{1}{(1-y q^i)(1-\frac{q^{-i-1}}{y})} &= \frac{q}{(1-q^{k+1} y)(1-\frac{q^{k}}{y})}  \frac{q^{2k+1}-1}{q-1} \, , \label{h0-identity} \\
\sum_{i=-k}^k \frac{1}{(1-y q^i)(1-\frac{q^{N-1-i}}{y})} &= \sum_{i=0}^{N-2} \frac{1}{(1-q^{k-i} y)(1-\frac{q^{k+(N-1)-i}}{y})} \frac{q^{2k+(N-1)-2 i}-1}{q^{N-1}-1} \, , \label{hN-identity} 
\end{align}
for $k \ge 0$ and $N \ge 2$.
\end{lemma}

\Proof{
Consider first \eqref{h0-identity}. We prove it by induction. Setting $k=0$ we see
\be
\frac{1}{1-y} \frac{1}{1-\frac{1}{q y}} = \frac{1}{1-q y} \frac{1}{1-\frac1y} q \, ,
\ee
which indeed holds.  Now we assume that \eqref{h0-identity} holds for a certain $k$ and we prove the identity for $k+1$. For $k+1$ we use the induction hypothesis to write the left hand side of \eqref{h0-identity} as
\begin{align}
\sum_{i=-k-1}^{k+1} \frac{1}{(1-y q^i)(1-\frac{q^{-i-1}}{y})} = & \frac{1}{1-q^{-k-1} y} \frac{1}{1-\frac{q^k}{y}} +
 \frac{q}{(1-q^{k+1} y)(1-\frac{q^{k}}{y})}  \frac{q^{2k+1}-1}{q-1} \cr & + 
 \frac{1}{1-q^{k+1} y} \frac{1}{1-\frac{q^{-k-2}}{y}} \cr = &
  \frac{q}{(1-q^{k+2} y)(1-\frac{q^{k+1}}{y})}  \frac{q^{2k+3}-1}{q-1} \, ,
\end{align}
which holds by direct computation.

Consider now \eqref{hN-identity}. Again we proceed by induction. For $k=0$ we have to prove the identity
\be
\frac{1}{(1-y)}\frac{1}{(1-\frac{q^{N-1}}{y})} =\sum_{i=0}^{N-2} \frac{1}{(1-q^{-i} y)} \frac{1}{(1-\frac{q^{N-1-i}}{y})} \frac{q^{N-1-2 i}-1}{q^{N-1}-1} \, ,
\ee
or equivalently
\be
\frac{q^{N-1}-1}{1-y}\frac{1}{1-\frac{q^{N-1}}{y}} =\sum_{i=0}^{N-2} \frac{q^{N-1-2 i}-1}{1-q^{-i} y} \frac{1}{1-\frac{q^{N-1-i}}{y}} \, .
\ee
This identity holds if and only if the following identity
\be
\sum_{i=1}^{N-2} \frac{q^{N-1-2 i}-1}{1-q^{-i} y} \frac{1}{1-\frac{q^{N-1-i}}{y}} = 0 \label{k0intermediate}
\ee
is true and only the $i=0$ term in the sum is non trivial. To prove this, let us consider separately the cases where $N$ is odd or even.

If $N$ is odd we can split the sum as
\be
\sum_{i=1}^{N-2} = \sum_{i=1}^{\frac{N-3}{2}} + \sum_{i=\frac{N-1}{2}} + \sum_{i=\frac{N+1}{2}}^{N-2}
\ee
The middle term is identically zero, since $q^{(N-1)-2 i}$ equals $1$ when $i=\frac{N-1}{2}$. To evaluate the last term, let us relabel $i = (N-1)-j$.  Then since
\begin{align}
 \frac{q^{N-1-2 i}-1}{1-q^{-i} y} \frac{1}{1-\frac{q^{N-1-i}}{y}} =  \frac{q^{N-1-2 ( (N-1)-j)}-1}{1-q^{-( (N-1)-j)} y} \frac{1}{1-\frac{q^{N-1-( (N-1)-j)}}{y}} = -  \frac{q^{N-1-2 j}-1}{1-q^{-j} y} \frac{1}{1-\frac{q^{N-1-j}}{y}} 
\end{align}
the first and third sums cancel exactly and \eqref{k0intermediate} is proven.

Consider now $N$ even. Now we simply split
\be
\sum_{i=1}^{N-2} = \sum_{i=1}^{\frac{N-2}{2}} +  \sum_{i=\frac{N-2}{2}+1}^{N-2} \, .
\ee
By using the same relabelling $i = (N-1)-j$ we see that again the sums cancel each other. Therefore the first step of the induction is proven.

Now let us assume that the identity \eqref{hN-identity} holds for a generic $k$. To simplify the exposition introduce the notation
\be
g_{k,N} (i) = \frac{1}{(1-q^{k-i} y)(1-\frac{q^{k+(N-1)-i}}{y})} \frac{q^{2k+(N-1)-2 i}-1}{q^{N-1}-1} \, .
\ee
Using the induction hypothesis we can write
\begin{align}
\sum_{i=-k-1}^{k+1} \frac{1}{(1-y q^i)(1-\frac{q^{N-i-1}}{y})} = \frac{1}{(1-y q^{-k-1})(1-\frac{q^{N+k}}{y})} + \sum_{i=0}^{N-2} g_{k,N} (i) +   \frac{1}{(1-y q^{k+1})(1-\frac{q^{N-2-k}}{y})} \, .
\end{align}
We have to prove that this is equivalent to $\sum_{i=0}^{N-2} g_{k+1,N} (i) $. However since in $g_{k,N} (i)$, $k$ and $i$ only appear in the combination $k-i$, it is easy to see that $\sum_{i=0}^{N-2} g_{k+1,N} (i) = \sum_{i=0}^{N-2} g_{k,N} (i-1) $. Therefore we have to prove that 
\be
\frac{1}{(1-y q^{-k-1})(1-\frac{q^{N+k}}{y})} + \sum_{i=0}^{N-2} g_{k,N} (i) +   \frac{1}{(1-y q^{k+1})(1-\frac{q^{N-2-k}}{y})} -  \sum_{i=0}^{N-2} g_{k,N} (i-1) = 0 \, .
\ee
It is easy to see that
\be
\sum_{i=0}^{N-2} \left( g_{k,N} (i) -g_{k,N} (i-1) \right) = g_{k,N} (N-2) - g_{k,N} (-1)
\ee
since the terms in the sum cancel pairwise. Then it is only a matter of brute force to show that the identity
\be
\frac{1}{(1-y q^{-k-1})(1-\frac{q^{N+k}}{y})}  + \frac{1}{(1-y q^{k+1})(1-\frac{q^{N-2-k}}{y})} + g_{k,N} (N-2) - g_{k,N} (-1) = 0
\ee
holds.
}

We now can prove our 

\begin{proposition}
We have that
\begin{align}
\tilde{s}_0 (\tau) & = \frac{q}{1-q} \, , \label{apps0} \\
\tilde{s}_{N} (\tau) & = \frac{N-1}{1-q^{N-1}} \, , \label{appsN}
\end{align}
where $N \ge 2$
\end{proposition}

\Proof{
By definition we have
\be
\tilde{s}_{M} (\tau) = \lim_{k\rightarrow \infty} \sum_{i=-k}^k \frac{1}{(1-y q^i)(1-\frac{q^{N-1-i}}{y})} \, ,
\ee
with $M=0$ or $M \ge 2$. By the results of the previous Lemma
\begin{align}
\tilde{s}_0 (\tau) & = \lim_{k \rightarrow \infty}  \frac{1}{(1-q^{k+1} y)(1-\frac{q^{k}}{y})} q \frac{q^{2k+1}-1}{q-1}  \, , \\
\tilde{s}_{N} (\tau) & =  \lim_{k \rightarrow \infty}  \sum_{i=0}^{N-2} \frac{1}{(1-q^{k-i} y)(1-\frac{q^{k+(N-1)-i}}{y})} \frac{q^{2k+(N-1)-2 i}-1}{q^{N-1}-1} \, .
\end{align}
To take the limit, recall that $q = \e^{2 \pi \ii \tau}$ is an analytic function of $\tau$ defined in the upper half plane. Therefore $\lim_{k\rightarrow \infty} q^k = 0 $ since the oscillatory part is bounded.

Then the result \eqref{apps0} follows immediately. To derive \eqref{appsN} we commute the limit with the finite sum and evaluate the limit at $N$ and $i$ fixed. Each limit gives the result $\frac{1}{1-q^{N-1}}$; evaluating the finite sum gives the result.
}

Finally we have
\begin{proposition}
We have the following identity
\be
s_1 (\tau , z) = \frac{\theta_1 (\tau , z)}{\eta(\tau)^3} \, 2 F_2^{(2)} (\tau) + \theta_1 (\tau ,z) \, \mu (\tau , z) \, .
\ee
\end{proposition}

\Proof{
From the definition of $s_1 (\tau , z)$ we have
\begin{align}
s_1 (\tau , z) = \frac{\theta_1 (\tau ,z)}{\eta^3 (\tau)} \sum_{i \in \zed} \frac{1}{(1-y q^i) (1 - \frac1y q^{-i})} = - \frac{\theta_1 (\tau ,z)}{\eta^3 (\tau)} \sum_{i \in \zed} \frac{y q^i}{(1 - y q^i)^2} \, .
\end{align}
We can write
\be
\sum_{i \in \zed} \frac{y q^i}{(1 - y q^i)^2} = \frac{y}{(1-y)^2} + \sum_{i=1}^\infty  \frac{y q^i}{(1 - y q^i)^2} + \sum_{i=1}^\infty  \frac{y q^{-i}}{(1 - y q^{-i})^2} \, .
\ee
By using the expansions
\be
\frac{x y}{(1 - x y)^2} = \sum_{n=1}^\infty n x^n y^n \, , \qquad \frac{x^{-1} y}{(1 - x^{-1} y)^2} = \sum_{n=1}^\infty n \frac{ x^n}{ y^n} \, ,
\ee
we find
\be
 \sum_{i=1}^\infty  \frac{y q^i}{(1 - y q^i)^2} + \sum_{i=1}^\infty  \frac{y q^{-i}}{(1 - y q^{-i})^2} = \sum_{i=1}^\infty \sum_{n=1}^\infty n \left( y^n + \frac{1}{y^n} \right) q^{n i} = \sum_{d=1}^\infty \sum_{d \vert k} k \left( y^k + \frac{1}{y^k} \right) q^{d} \, .
\ee
Putting all together we see
\be
s_1 (\tau , z) = - \frac{\theta_1 (\tau ,z)}{\eta^3 (\tau)}  \left( 
\frac{y}{(1-y)^2} + \sum_{d=1}^\infty \sum_{d \vert k} k \left( y^k + \frac{1}{y^k} \right) q^{d}
\right)
\ee
and the Proposition follows from the identity \eqref{vertmu}
}

\section{The Eisenstein series $E_2 (\tau)$ and the cyclotomic polynomials} \label{cyclo}

To have a different perspective on the rational function in \eqref{HEll2}, let us consider a truncated expansion of the Eisenstein series $E_2 (\tau)$ in terms of cyclotomic polynomials. Recall that the cyclotomic polynomials are defined as
\be
\Phi_n (x) = \prod_{\stackrel{1 \le k \le n}{\mathrm{gcd}(k,n)=1}} \left( x - \e^{2 \pi \ii \frac{k}{n}} \right) \, .
\ee
In particular one has
\be \label{cycloidentity}
x^n - 1 = \prod_{d \vert n} \Phi_d (x) \, .
\ee
Consider now the following truncation of the Eisenstein series
\be
E_2^{(N)} (\tau) = 1 - 24 \sum_{n=1}^N \frac{n q^n}{1 - q^n} = 1 - 24 \sum_{n=1}^N \frac{n}{1-q^n} + 24 \sum_{n=1}^N n \, .
\ee
Using \eqref{cycloidentity} we can write
\be
- \frac{n}{1-q^n} = n \prod_{d \vert n} \frac{1}{\Phi_d (q)} \, .
\ee
It is easy to see that
\be \label{cyclorem}
 n \prod_{d \vert n} \frac{1}{\Phi_d (q)} = \sum_{d \vert n} \frac{\mathrm{Reminder} \left( q \, \Phi'_d (q)  ; \Phi_d (q) \right)}{\Phi_d (q)} \, ,
\ee
expressed in terms of the reminder of the division of the polynomial $ q \, \Phi'_d (q)$ by $ \Phi_d (q)$. Indeed by taking the logarithmic derivative of \eqref{cycloidentity} one finds
\be
\frac{n \, q^n}{\prod_{d \vert n} \Phi_d (q) } = \sum_{d \vert n} \frac{q \, \Phi'_d (q) }{\Phi_d (q)} \, ,
\ee
which implies, by adding and subtracting 1 to the numerator on the left hand side 
\be \label{cyclorem2}
\frac{n }{\prod_{d \vert n} \Phi_d (q) } = - n +  \sum_{d \vert n} \frac{q \, \Phi'_d (q) }{\Phi_d (q)} \, .
\ee
The result follows from the fact that
\be \label{cyclorem3}
\frac{q \, \Phi'_d (q) }{\Phi_d (q)} = \phi (d) + \frac{\mathrm{Reminder} \left( q \, \Phi'_d (q)  ; \Phi_d (q) \right)}{\Phi_d (q)}
\ee 
which for example can be seen directly using the long polynomial division algorithm. Here $\phi(d)$ denotes the degree of $\Phi_d (q)$. Finally \eqref{cyclorem} follows by substituting \eqref{cyclorem3} into \eqref{cyclorem2} and using $\sum_{d \vert n} \phi(d) = n$.

Therefore we can write
\begin{align}
E_2^{(N)} (\tau) &= 1 + 24  \sum_{n=1}^N \sum_{d \vert n} \frac{\mathrm{Reminder} \left( q \, \Phi'_d (q)  ; \Phi_d (q) \right)}{\Phi_d (q)} + 24 \sum_{n=1}^N n \\
& = 1 + 24  \sum_{n=1}^N  r_N (n)  \frac{\mathrm{Reminder} \left( q \, \Phi'_n (q)  ; \Phi_n (q) \right)}{\Phi_n (q)} + 24 \sum_{n=1}^N n \, , \nonumber
\end{align}
where the coefficients $r_N (n)$ count the number of times an integer $n$ appears as a divisor of the set of integers $\{ 1 , \dots , N \}$
\be
r_N (n) = \mathrm{Coeff} \left( \sum_{k=1}^N \sum_{i \vert k} x^i \ ; \
x^n
\right) \, .
\ee
Since we can write
\be
\frac{q^4 + 2 q^3 + 6 q^2 + 2 q + 1}{(q-1) (q+1) (q^2+ q + 1)} = -1 -2 \frac{1}{\Phi_1 (q)} -2 \frac{1}{\Phi_2 (q)} +  \frac{\mathrm{Reminder} \left(q \, \Phi'_3 (q) ; \Phi_3 (q)  \right)}{\Phi_3 (q)} \, ,
\ee
the rational function can be thought of as a shift of the first coefficients of the Eisenstein series.


\begin{thebibliography}{99}

%\cite{Alexandrov:2020qpb}
\bibitem{Alexandrov:2020qpb}
S.~Alexandrov and S.~Nampuri,
``Refinement and modularity of immortal dyons,''
JHEP \textbf{01} (2021), 147
%doi:10.1007/JHEP01(2021)147
[arXiv:2009.01172 [hep-th]].
%0 citations counted in INSPIRE as of 05 Mar 2023

%\cite{Alexandrov:2019rth}
\bibitem{Alexandrov:2019rth}
S.~Alexandrov, J.~Manschot and B.~Pioline,
``S-duality and refined BPS indices,''
Commun. Math. Phys. \textbf{380} (2020) no.2, 755-810
%doi:10.1007/s00220-020-03854-6
[arXiv:1910.03098 [hep-th]].
%24 citations counted in INSPIRE as of 05 Mar 2023

%\cite{Anagiannis:2018jqf}
\bibitem{Anagiannis:2018jqf}
V.~Anagiannis and M.~C.~N.~Cheng,
``TASI Lectures on Moonshine,''
PoS \textbf{TASI2017} (2018), 010
%doi:10.22323/1.305.0010
[arXiv:1807.00723 [hep-th]].
%13 citations counted in INSPIRE as of 26 Mar 2023

%\cite{Antoniadis:2006mr}
\bibitem{Antoniadis:2006mr}
I.~Antoniadis, S.~Hohenegger and K.~S.~Narain,
``N=4 Topological Amplitudes and String Effective Action,''
Nucl. Phys. B \textbf{771} (2007), 40-92
%doi:10.1016/j.nuclphysb.2007.02.011
[arXiv:hep-th/0610258 [hep-th]].
%21 citations counted in INSPIRE as of 02 Dec 2022

%\cite{Benjamin:2016pil}
\bibitem{Benjamin:2016pil}
N.~Benjamin,
``A Refined Count of BPS States in the D1/D5 System,''
JHEP \textbf{06} (2017), 028
%doi:10.1007/JHEP06(2017)028
[arXiv:1610.07607 [hep-th]].
%13 citations counted in INSPIRE as of 18 Jan 2023

%\cite{Benjamin:2017xen}
\bibitem{Benjamin:2017xen}
N.~Benjamin, S.~Kachru and A.~Tripathy,
``Counting spinning dyons in maximal supergravity: The Hodge-elliptic genus for tori,''
Lett. Math. Phys. \textbf{107} (2017) no.11, 2081-2092
%doi:10.1007/s11005-017-0981-8
[arXiv:1704.05423 [hep-th]].
%2 citations counted in INSPIRE as of 18 Jan 2023

%\cite{Benjamin:2017rnd}
\bibitem{Benjamin:2017rnd}
N.~Benjamin and S.~M.~Harrison,
``Symmetries of the refined D1/D5 BPS spectrum,''
JHEP \textbf{11} (2017), 091
%doi:10.1007/JHEP11(2017)091
[arXiv:1708.02244 [hep-th]].
%5 citations counted in INSPIRE as of 18 Jan 2023

%\cite{Bryan:2015uva}
\bibitem{Bryan:2015uva}
  J.~Bryan,
  ``The Donaldson-Thomas theory of $K3\times E$ via the topological vertex,'' in Geometry of Moduli, (eds. Christophersen, J. A. and Ranestad, K.) Abel Symposia, 14 (Springer, 2018).
  arXiv:1504.02920 [math.AG].
  %%CITATION = ARXIV:1504.02920;%%
  %7 citations counted in INSPIRE as of 04 Jul 2019

%\cite{Bryan:2016bse}
\bibitem{Bryan:2016bse}
J.~Bryan and M.~Kool,
``Donaldson\textendash{}thomas Invariants of Local Elliptic Surfaces via the Topological Vertex,''
Forum Math. Sigma \textbf{7} (2019), e7
%doi:10.1017/fms.2019.1
[arXiv:1608.07369 [math.AG]].
%5 citations counted in INSPIRE as of 25 Mar 2023

%\cite{Bryan:2016gfy}
\bibitem{Bryan:2016gfy}
J.~Bryan, M.~Kool and B.~Young,
``Trace Identities for the Topological Vertex,''
Selecta Math. \textbf{24} (2018), 1527-1548
%doi:10.1007/s00029-017-0302-1
[arXiv:1603.05271 [math.CO]].
%4 citations counted in INSPIRE as of 25 Mar 2023

%\cite{Bryan:2018nlv}
\bibitem{Bryan:2018nlv}
J.~Bryan and G.~Oberdieck,
``CHL Calabi\textendash{}Yau threefolds: curve counting, Mathieu moonshine and Siegel modular forms,''
Commun. Num. Theor. Phys. \textbf{14} (2020) no.4, 785-862
%doi:10.4310/CNTP.2020.v14.n4.a3
[arXiv:1811.06102 [math.AG]].
%7 citations counted in INSPIRE as of 25 Mar 2023

%\cite{Chattopadhyaya:2017ews}
\bibitem{Chattopadhyaya:2017ews}
A.~Chattopadhyaya and J.~R.~David,
``Dyon degeneracies from Mathieu moonshine symmetry,''
Phys. Rev. D \textbf{96} (2017) no.8, 086020
%doi:10.1103/PhysRevD.96.086020
[arXiv:1704.00434 [hep-th]].
%12 citations counted in INSPIRE as of 25 Mar 2023


%\cite{Cheng:2015kha}
\bibitem{Cheng:2015kha}
M.~C.~N.~Cheng, J.~F.~R.~Duncan, S.~M.~Harrison and S.~Kachru,
``Equivariant K3 Invariants,''
Commun. Num. Theor. Phys. \textbf{11} (2017), 41-72
%doi:10.4310/CNTP.2017.v11.n1.a2
[arXiv:1508.02047 [hep-th]].
%17 citations counted in INSPIRE as of 26 Mar 2023

%\cite{Cheng:2016org}
\bibitem{Cheng:2016org}
M.~C.~N.~Cheng, S.~M.~Harrison, R.~Volpato and M.~Zimet,
``K3 String Theory, Lattices and Moonshine,''
[arXiv:1612.04404 [hep-th]].
%30 citations counted in INSPIRE as of 25 Mar 2023

%\cite{Cirafici:2021yda}
\bibitem{Cirafici:2021yda}
M.~Cirafici,
``On the M2-Brane Index on Noncommutative Crepant Resolutions,''
 \textit{Lett Math Phys} \textbf{112}, 88 (2022).
[arXiv:2111.01102 [hep-th]].
%0 citations counted in INSPIRE as of 31 Oct 2022

\bibitem{mathfile}
M.~Cirafici, supporting \textsc{mathematica} file, available at  \url{https://cirafici.dmg.units.it/HEsupporting.nb} or as an ancillary file in the arXiv submission.


%\cite{Creutzig:2013mqa}
\bibitem{Creutzig:2013mqa}
T.~Creutzig and G.~H\"ohn,
%``Mathieu Moonshine and the Geometry of K3 Surfaces,''
Commun. Num. Theor. Phys. \textbf{08} (2014), 295-328
%doi:10.4310/CNTP.2014.v8.n2.a3
[arXiv:1309.2671 [math.QA]].
%16 citations counted in INSPIRE as of 07 Oct 2022

%\cite{Dabholkar:2012nd}
\bibitem{Dabholkar:2012nd}
A.~Dabholkar, S.~Murthy and D.~Zagier,
``Quantum Black Holes, Wall Crossing, and Mock Modular Forms,''
[arXiv:1208.4074 [hep-th]].
%186 citations counted in INSPIRE as of 05 Oct 2022

%\cite{David:2006ud}
\bibitem{David:2006ud}
J.~R.~David, D.~P.~Jatkar and A.~Sen,
``Dyon spectrum in generic N=4 supersymmetric Z(N) orbifolds,''
JHEP \textbf{01} (2007), 016
%doi:10.1088/1126-6708/2007/01/016
[arXiv:hep-th/0609109 [hep-th]].
%91 citations counted in INSPIRE as of 25 Mar 2023

%\cite{Dijkgraaf:1996xw}
\bibitem{Dijkgraaf:1996xw}
R.~Dijkgraaf, G.~W.~Moore, E.~P.~Verlinde and H.~L.~Verlinde,
``Elliptic genera of symmetric products and second quantized strings,''
Commun. Math. Phys. \textbf{185} (1997), 197-209
%doi:10.1007/s002200050087
[arXiv:hep-th/9608096 [hep-th]].
%355 citations counted in INSPIRE as of 06 Oct 2022

%\cite{Duncan:2015xoa}
\bibitem{Duncan:2015xoa}
J.~F.~R.~Duncan and S.~Mack-Crane,
``Derived Equivalences of K3 Surfaces and Twined Elliptic Genera,''
[arXiv:1506.06198 [math.RT]].
%18 citations counted in INSPIRE as of 06 Oct 2022

%\cite{Eguchi:2010fg}
\bibitem{Eguchi:2010fg}
T.~Eguchi and K.~Hikami,
``Note on twisted elliptic genus of $K3$ surface,''
Phys. Lett. B \textbf{694} (2011), 446-455
%doi:10.1016/j.physletb.2010.10.017
[arXiv:1008.4924 [hep-th]].
%103 citations counted in INSPIRE as of 25 Mar 2023

%\cite{Eguchi:2010ej}
\bibitem{Eguchi:2010ej}
  T.~Eguchi, H.~Ooguri and Y.~Tachikawa,
  ``Notes on the K3 Surface and the Mathieu group $M_{24}$,''
  Exper.\ Math.\  {\bf 20} (2011) 91
%  doi:10.1080/10586458.2011.544585
  [arXiv:1004.0956 [hep-th]].
  %%CITATION = doi:10.1080/10586458.2011.544585;%%
  %139 citations counted in INSPIRE as of 16 Jul 2019
  
%\cite{Eguchi:1987sm}
\bibitem{Eguchi:1987sm}
T.~Eguchi and A.~Taormina,
``Unitary Representations of $N=4$ Superconformal Algebra,''
Phys. Lett. B \textbf{196} (1987), 75
%doi:10.1016/0370-2693(87)91679-0
%106 citations counted in INSPIRE as of 10 Oct 2022

%\cite{Eguchi:1987wf}
\bibitem{Eguchi:1987wf}
T.~Eguchi and A.~Taormina,
``Character Formulas for the $N=4$ Superconformal Algebra,''
Phys. Lett. B \textbf{200} (1988), 315
%doi:10.1016/0370-2693(88)90778-2
%136 citations counted in INSPIRE as of 10 Oct 2022

%\cite{Eguchi:1988af}
\bibitem{Eguchi:1988af}
T.~Eguchi and A.~Taormina,
``On the Unitary Representations of $N=2$ and $N=4$ Superconformal Algebras,''
Phys. Lett. B \textbf{210} (1988), 125-132
%doi:10.1016/0370-2693(88)90360-7
%117 citations counted in INSPIRE as of 10 Oct 2022

%\cite{Harvey:2017xdt}
\bibitem{Harvey:2017xdt}
J.~A.~Harvey and G.~W.~Moore,
``Conway Subgroup Symmetric Compactifications of Heterotic String,''
J. Phys. A \textbf{51} (2018) no.35, 354001
%doi:10.1088/1751-8121/aac9d1
[arXiv:1712.07986 [hep-th]].
%12 citations counted in INSPIRE as of 19 Mar 2023

%\cite{Gaberdiel:2010ch}
\bibitem{Gaberdiel:2010ch}
M.~R.~Gaberdiel, S.~Hohenegger and R.~Volpato,
``Mathieu twining characters for K3,''
JHEP \textbf{09} (2010), 058
%doi:10.1007/JHEP09(2010)058
[arXiv:1006.0221 [hep-th]].
%105 citations counted in INSPIRE as of 25 Mar 2023

%\cite{Gaberdiel:2011fg}
\bibitem{Gaberdiel:2011fg}
M.~R.~Gaberdiel, S.~Hohenegger and R.~Volpato,
``Symmetries of K3 sigma models,''
Commun. Num. Theor. Phys. \textbf{6} (2012), 1-50
%doi:10.4310/CNTP.2012.v6.n1.a1
[arXiv:1106.4315 [hep-th]].
%86 citations counted in INSPIRE as of 25 Mar 2023

%\cite{Gaberdiel:2012gf}
\bibitem{Gaberdiel:2012gf}
M.~R.~Gaberdiel, D.~Persson, H.~Ronellenfitsch and R.~Volpato,
``Generalized Mathieu Moonshine,''
Commun. Num. Theor Phys. \textbf{07} (2013), 145-223
%doi:10.4310/CNTP.2013.v7.n1.a5
[arXiv:1211.7074 [hep-th]].
%51 citations counted in INSPIRE as of 25 Mar 2023

\bibitem{kac}
V.~G.~Kac, M.~Wakimoto, Integrable Highest Weight Modules over Affine Superalgebras and Number Theory (1994). In: Brylkinski, JL, Brylinski, R. Guillemin, V.,Kac, V. (eds) Lie Theory and Geometry. Progress in Mathematics, vol 123. Birkhauser, Boston, MA.

%\cite{Kachru:2016igs}
\bibitem{Kachru:2016igs}
  S.~Kachru and A.~Tripathy,
  ``The Hodge-elliptic genus, spinning BPS states, and black holes,''
  Commun.\ Math.\ Phys.\  {\bf 355} (2017) no.1,  245
%  doi:10.1007/s00220-017-2910-1
  [arXiv:1609.02158 [hep-th]].
  %%CITATION = doi:10.1007/s00220-017-2910-1;%%
  %9 citations counted in INSPIRE as of 04 Jul 2019

%\cite{Kachru:2017yda}
\bibitem{Kachru:2017yda}
S.~Kachru and A.~Tripathy,
``BPS jumping loci and special cycles,''
[arXiv:1703.00455 [hep-th]].
%10 citations counted in INSPIRE as of 18 Jan 2023  
  
%\cite{Kachru:2017zur}
\bibitem{Kachru:2017zur}
S.~Kachru and A.~Tripathy,
``BPS jumping loci are automorphic,''
Commun. Math. Phys. \textbf{360} (2018) no.3, 919-933
%doi:10.1007/s00220-018-3090-3
[arXiv:1706.02706 [hep-th]].
%5 citations counted in INSPIRE as of 18 Jan 2023  
  
 %\cite{Kapustin:2005pt}
\bibitem{Kapustin:2005pt}
A.~Kapustin,
``Chiral de Rham complex and the half-twisted sigma-model,''
[arXiv:hep-th/0504074 [hep-th]].
%72 citations counted in INSPIRE as of 19 Mar 2023 
    
 
%\cite{Katz:2014uaa}
\bibitem{Katz:2014uaa}
  S.~Katz, A.~Klemm and R.~Pandharipande,
  ``On the motivic stable pairs invariants of K3 surfaces,''
  arXiv:1407.3181 [math.AG].
  %%CITATION = ARXIV:1407.3181;%%
  %16 citations counted in INSPIRE as of 04 Jul 2019

%\cite{Katz:1999xq}
\bibitem{Katz:1999xq}
S.~H.~Katz, A.~Klemm and C.~Vafa,
``M theory, topological strings and spinning black holes,''
Adv. Theor. Math. Phys. \textbf{3} (1999), 1445-1537
%doi:10.4310/ATMP.1999.v3.n5.a6
[arXiv:hep-th/9910181 [hep-th]].
%205 citations counted in INSPIRE as of 17 Mar 2023

%\cite{Kawai:1997em}
\bibitem{Kawai:1997em}
T.~Kawai,
``K3 surfaces, Igusa cusp form and string theory,''
[arXiv:hep-th/9710016 [hep-th]].
%19 citations counted in INSPIRE as of 26 Mar 2023

%\cite{Kawai:1993jk}
\bibitem{Kawai:1993jk}
T.~Kawai, Y.~Yamada and S.~K.~Yang,
``Elliptic genera and N=2 superconformal field theory,''
Nucl. Phys. B \textbf{414} (1994), 191-212
%doi:10.1016/0550-3213(94)90428-6
[arXiv:hep-th/9306096 [hep-th]].
%171 citations counted in INSPIRE as of 26 Mar 2023

\bibitem{chiraldR}
Malikov, F., Schechtman, V. Vaintrob, A. \textit{Chiral de Rham Complex},  Comm Math Phys \textbf{204}, 439-473 (1999).

%\bibitem{felix}
%G.~W.~Moore, \textit{Felix Klein Lectures}, Hausdorff Mathematical Institute, Bonn 1-11 October 2012.

\bibitem{membranes}
N.~Nekrasov, A.~Okounkov,  \textit{Membranes and Sheaves}. Algebraic Geometry \textbf{3}(3), 320-369 (2016). arXiv:1404.2323 [math.AG]

\bibitem{OP}
G.~Oberdieck, R.~Pandharipande, Curve counting on $K3 \times E$, the Igusa cusp form $\chi_{10}$, and descendent integration, in K3 surfaces and their moduli, C.~Faber, G.~Farkas,
and G. van der Geer, eds., Birkhauser Prog. in Math. 315 (2016), 245-278.

%\cite{Oberdieck:2017zit}
\bibitem{Oberdieck:2017zit}
G.~Oberdieck and A.~Pixton,
``Holomorphic anomaly equations and the Igusa cusp form conjecture,''
Invent. Math. \textbf{213} (2018), 507
%doi:10.1007/s00222-018-0794-0
[arXiv:1706.10100 [math.AG]].
%23 citations counted in INSPIRE as of 17 Mar 2023

%\cite{Sen:2007qy}
\bibitem{Sen:2007qy}
A.~Sen,
``Black Hole Entropy Function, Attractors and Precision Counting of Microstates,''
Gen. Rel. Grav. \textbf{40} (2008), 2249-2431
%doi:10.1007/s10714-008-0626-4
[arXiv:0708.1270 [hep-th]].
%400 citations counted in INSPIRE as of 26 Mar 2023

%\cite{Sen:2012hv}
\bibitem{Sen:2012hv}
A.~Sen,
``BPS Spectrum, Indices and Wall Crossing in N=4 Supersymmetric Yang-Mills Theories,''
JHEP \textbf{06} (2012), 164
%doi:10.1007/JHEP06(2012)164
[arXiv:1203.4889 [hep-th]].
%8 citations counted in INSPIRE as of 06 Mar 2023

%\cite{Wendland:2017eiw}
\bibitem{Wendland:2017eiw}
K.~Wendland,
``Hodge-Elliptic Genera and How They Govern K3 Theories,''
Commun. Math. Phys. \textbf{368} (2019) no.1, 187-221
%[arXiv:1705.09904 [hep-th]].
%11 citations counted in INSPIRE as of 12 Jan 2023

%\cite{Witten:1986bf}
\bibitem{Witten:1986bf}
E.~Witten,
``Elliptic Genera and Quantum Field Theory,''
Commun. Math. Phys. \textbf{109} (1987), 525
%doi:10.1007/BF01208956
%295 citations counted in INSPIRE as of 26 Mar 2023

\end{thebibliography}
\end{document}